\documentclass[10pt]{article}
\usepackage{graphicx}
\usepackage{amssymb}
\usepackage{epstopdf}
\usepackage{color}
\usepackage{amsmath}
\usepackage{amsfonts}
\usepackage{dsfont}
\usepackage{booktabs}

\begin{document}

\title{Calibration of optimal execution of financial transactions in the presence of transient market impact}
\author{Enzo Busseti\,$^{\textrm{a}, \textrm{b}}$ and Fabrizio Lillo\,$^{\textrm{a}, \textrm{c}, \textrm{d}}$\footnote{We acknowledge the Linear Quantitative Research team at J.P. Morgan for providing support, insights, advices and data. We thank Jean-Philippe Bouchaud, Gianbiagio Curato, Daniel Nehren, Xavier Abdobal, Alex Alekseev, Oleg Bogoslavskiy, Junfeng  Li, Sami  Haj Slimane for useful discussion. This work is partially supported by the grant SNS11LILLB ``Price formation, agent's heterogeneity, and market efficiency"}}
\date{}
\maketitle
\small
\begin{center}
  $^\textrm{a}$~\emph{Scuola Normale Superiore di Pisa, Piazza dei Cavalieri 7, I-56126 Pisa, Italy}\\
  $^\textrm{b}$~\emph{J.P. Morgan, Linear Quantitative Research, 10 Aldermanbury, London EC2V 7RF, UK}\\
  $^\textrm{c}$~\emph{Dipartimento di Fisica, viale delle Scienze I-90128, Palermo, Italy}\\
  $^\textrm{d}$~\emph{Santa Fe Institute, 1399 Hyde Park Road, Santa Fe NM 87501, USA}\\ 
\end{center}
\normalsize

\begin{abstract}
Trading large volumes of a financial asset in order driven markets requires the use of algorithmic execution dividing the volume in many transactions in order to minimize costs due to market impact. A proper design of an optimal execution strategy strongly depends on a careful modeling of market impact, i.e. how the price reacts to trades.  In this paper we consider a recently introduced market impact model (Bouchaud {\it et al.}, 2004, \cite{bouchaud2004fluctuations}), which has the property of describing both the volume and the temporal dependence of price change due to trading. We show how this model can be used to describe price impact also in aggregated trade time or in real time.  We then solve analytically and calibrate with real data the optimal execution problem both for risk neutral and for risk averse investors and we derive an efficient frontier of optimal execution.  When we include spread costs the problem must be solved numerically and we show that the introduction of such costs regularizes the solution. 
\end{abstract}

\newpage

\tableofcontents 

\section{Introduction}

Optimal execution of portfolio transactions is receiving a growing attention in the recent years. Due in part to the increasing automation of the trading process, optimal trading is becoming also an industrial standard, not only for very large investors. The problem is generically stated in the following way. An investor wants to trade (buy or sell) a given number of shares and wants to minimize cost. Execution cost has many components, such as fees, taxes, investment delays, market impact, timing risk, and opportunity cost \cite{kissell2003optimal}. These components depend on the execution algorithm and in many cases are stochastic variables. This means that their value depends on the specific realization of the trading process. Moreover stochastic costs are associated to a risk, which is related to the uncertainty of the cost. As usual in finance, this fact entails a risk-reward (in this case, cost) tradeoff, which can be solved by choosing an appropriate level of risk aversion. 

In modern electronic markets, the problem of optimal execution can be a multi time scale problem. It is not unlikely that a portfolio manager wants to trade an order whose size requires an execution over multiple days. In this case we can imagine a three scale decomposition of the execution. First, the portfolio manager decides how to split the order across the different days, possibly dividing it also among different brokers to minimize information leakage. In this case there are specific risks, such as those due to overnight returns. Then, for each day the trader divides the day in ``macroscopic" intervals, for example $5$ or $15$ minutes, and decides how much to trade in each of these intervals. Lastly, one has to decide how to trade in each interval, specifying the type of orders used (e.g. limit versus market orders) and the strategy to follow (for example, when to cross the spread if the price moves in an adverse direction). Although this division is very schematic and the optimization is not necessarily segmented rigidly in this way, this scheme gives an idea of different types of optimization that can be used as building blocks to simplify the whole optimization problem.

In this paper, we will focus on the second level of trading optimization, namely we will assume that a investor wants to trade a position throughout the day and that she considers the day divided in intervals of several minutes, searching the optimal way of splitting the order among the intervals. Moreover we will focus our attention to price impact cost only. However, since we will assume that the investor trades by using market orders, bid ask spread costs cannot be neglected, and in fact we will see that they play a very important role in the solution we obtain. 

There is a vast literature on optimal execution of portfolio transactions. The seminal paper of Bertsimas and Lo \cite{bertsimas1998optimal} considered the optimal execution problem by modeling a linear and permanent impact and by minimizing the expected cost. Not surprisingly they found that the optimal solution is to ``slice and dice" the order, i.e. to divide it in equal parts and trade it at constant rate. More recently, Almgren and Chriss \cite{almgren2001optimal} introduced two important innovations. First beside the (linear) permanent impact, they considered a temporary component of the impact, which takes into account the fact that trading with market orders implies a liquidity cost. The second innovation was to include risk aversion in the optimization. Since traders can be risk averse, they will minimize expected cost for a given level of risk (or, equivalently, minimize risk for a given level of expected cost). Therefore Almgren and Chris minimized an objective function which is a linear combination of expected cost and risk (quantified as the variance of cost), and the coefficient in front of the risk term is a measure of risk aversion (see Section 2.1). 

A key aspect in modeling optimal execution is market impact. Roughly speaking, market impact is the relation between signed trade volume and price change \cite{bouchaud2009markets}. In general this relation is complicated, because it is a function of the type of order, the sign of the trade, the volume, the available liquidity, etc.. Several empirical and theoretical studies have considered the properties of market impact \cite{hopman2002supply,lillo2003master,bouchaud2004fluctuations,farmer2004really}.  The two approaches to optimal execution mentioned above \cite{bertsimas1998optimal,almgren2001optimal}, and many others published in the literature, assume (i) that contemporaneous impact is linear in volume and (ii) that impact has no transient component, i.e. the trades made at a given time have no effect on the {\it returns} caused by other trades made at a subsequent time. Empirical research indicates that these two assumptions are not consistent with real data. The impact of individual transactions is a strongly concave function of the volume \cite{lillo2003master,potters2003more}. More importantly, it has been recently shown \cite{bouchaud2004fluctuations,bouchaud2009markets} that market impact has a significant transient component. We should stress that the transient component of impact is strongly related to market efficiency. In fact order flow is highly persistent in time  \cite{lillo2004long} and, if the impact were fixed and permanent, also price movements would be strongly autocorrelated in time. We note that it has been shown empirically \cite{imon} that the persistence of the order flow is mainly determined by order splitting, i.e. optimal execution.    

In this paper we will consider the problem of calibrating the optimal execution of financial transactions in the presence of transient market impact. From a theoretical point of view this problem has been sketched in  \cite{bouchaud2009markets} and considered recently in \cite{alfonsi2009order}, where it has been discussed in the framework of the impossibility of price manipulation. Some of these theoretical solutions will be re-derived here. Moreover our optimal solution will be compared with the one proposed by Obizhaeva and Wang \cite{obizhaeva2006optimal} as a solution of a model that takes into account a specific mechanism of spread resilience.
Then we will consider the more realistic case where one considers the bid ask spread cost and we will show how the inclusion of spread cost regularizes the solution. Apart from theoretical analysis, the paper will be focused on the calibration on real data. To this end we will show how to  define and calibrate a transient impact model when we consider a macroscopic (i.e. non tick by tick) time measure. In this work we consider real time and aggregated trade time.

The paper is divided as follows. In Section 2 we fix the notation and we review the optimal execution problem. In Section 3 we present the dataset we used. In Section 4 we review the propagator model in trade time, as introduced in Ref. \cite{bouchaud2004fluctuations}. Section 5 discusses how to extend and calibrate the propagator model in real time (the Appendix discusses the calibration in aggregated trade time). In Section 6 we discuss the solutions of the optimal execution algorithms and in Section 7 we calibrate them to the real data. Finally, in Section 8 we draw some conclusions.

\section{Optimal execution problem}
\label{sec:statement_of_the_problem}

In this section we review the problem of optimal execution and we introduce the definitions that we will use in the rest of this work.
We  define two important quantities, namely the \emph{trading schedule} and the \emph{execution cost}. 

Consider an investor who wants to trade a block of $X$ shares of a stock. If $X$ is positive (negative) the investor wants to buy (sell). 
We will consider execution programs that make use only of market orders. In this way there is certainty that the order will be executed, but the bid ask spread cost might be high.
We assume that the investor wants to execute her trade in a predetermined time horizon $T$, as, for example, one trading day.
We divide $T$ into $N$ intervals of length $\tau = \frac{T}{N}$, and call $t_k = k\tau$ ($k = 0,1,...,N-1$) the initial times of each interval. The value of $N$ is predetermined by the investor.  

We define a \emph{trading schedule} as the $N$ dimensional vector $\mathbf v=(v_0, v_1, ..., v_{N-1})^T$, where $T$ denotes the transpose. Each element $v_k $  is the signed volume traded between time $t_{k}$ and $t_{k+1}$. Conventionally, the sign is positive (negative) if the agent buys (sells).  We clearly have that

\begin{equation}
X = \sum_{j=0}^{N-1}v_j. 
\end{equation}

A \emph{trading strategy} is now easily defined as a rule for determining the trading schedule $v_k$ with the information available at time $t_{k-1}$. Note that at this point we are not imposing that the sign of each $v_k$ is equal to the sign of $X$. In fact, a buy trading strategy can in principle include sell transactions.

Each trading schedule is associated with an execution cost. In order to define cost, we need to introduce a notation for price. The mid price at the start of interval $k$ is denoted by $P_k$ and its logarithm is $p_k=\ln P_k$. Note that the average price at which a share is traded at time $k$ is not $P_k$. In fact commissions fees and, most notably, bid-ask spread raise this price for buy trades or lower it for sell trades. We therefore define the series $\tilde{P_k}$ which represent the effective prices at which shares are actually traded at every step. This is similar to what is done in Almgren and Chriss approach \cite{almgren2001optimal} (see also below). The precise relationship between  $\tilde{P_k}$ and $P_k$ will be clarified later on. 

We define the execution cost $C (\mathbf v)$ as the difference between the total money payed or received during the execution (equal to the sum, over all intervals, of the volume traded $v_k$ and the corresponding effective price $\tilde{P_k}$) and the initial market value of the position. In formulas it is

\begin{equation}
 C (\mathbf v) = \sum_{k=0}^{N-1} v_k  \tilde{P_k} - X P_0=\sum_{k=0}^{N-1} v_k (\tilde{P}_k - P_0).
\label{eq:cost_of_trajectory}
\end{equation}
Execution cost so defined is equivalent to the implementation shortfall introduced in Ref. \cite{perold1998implementation}. For a buy trade it is the difference between the price payed and the price that would have been payed in an infinitely liquid market. We note that execution costs needs not to be positive, i.e. one may have negative costs (gains) from the execution. For example, if during a buying schedule the stock price declines substantially, $C (\mathbf v)$ could be negative because the total money paid is less than the initial market value of the position.

It is important to stress that, since price is a stochastic process, whose properties might be affected by the trading schedule, also execution cost is a stochastic process. Different realizations of the price process under the same trading schedule give different execution costs. For this reason it is important to characterize its expected value and its variance,
\begin{equation}
E\left[C (\mathbf v)\right], ~~~~~~~~~Var\left[C (\mathbf v)\right] =  E \left[ {\left(C (\mathbf v) - [C (\mathbf v)] \right)}^2\right].
\end{equation}
The first quantity measures the expected cost, while the variance quantifies the risk associated to the execution.

\subsection{The Almgren Chriss execution scheme}

The problem of optimal execution, as we formulated it, has been the focus of many studies.  One particularly influential work is the 2001 paper by Almgren and Chriss \cite{almgren2001optimal}. They assume that the price of the stock at step $k$ is equal to the previous price plus a linear market impact term and a random shock:

\begin{equation}
P_k = P_{k-1}  + \theta v_{k-1} + \eta_{k-1}\;\;\;\;~~ \eta \sim \mbox{IID}(0, \sigma) .
\label{priceDyn}
\end{equation}
This specification for the price impact is also used in Ref. \cite{bertsimas1998optimal}.

The first  innovations in Almgren and Chriss is that they consider the effective price $\tilde{P_k}$ and they model it as
\begin{equation}
  \tilde{P_k} = P_{k}  + \rho  v_{k} + \mbox{sign}(v_k)\cdot  S/2,
  \label{eq:almgren_chriss_effective_price}
\end{equation}
where the term $\mbox{sign}(v_k)\cdot S/2$ is the contribution from the bid ask spread $S$ and $ \rho v_{k}$ represents a linear {\it temporary impact}. In the terminology of \cite{almgren2001optimal}, the temporary impact accounts for the resilience of the limit orders in the book, which relaxes back to the steady state after a trade-induced price movement.

The equation \ref{eq:cost_of_trajectory} for the execution costs becomes

\begin{equation}
 C (\mathbf v) = \sum_{k=0}^{N-1} v_k  \tilde{P_k} - X P_0 =  
 \sum_{k=0}^{N-1} \left( \eta_k + \theta v_k  \right) \sum_{j = k+1}^{N-1} v_j 
 + \sum_{k=0}^{N-1} \left(   \mbox{sign}(v_k) S/2 +  \rho v_{k} \right) v_k
\label{trajectoryRevenueTemporaryImpact}
\end{equation}
and the expected value of the costs is

\begin{equation}
E \left[C (\mathbf v)\right] = \frac{\theta}{2} X^2 + (\rho - \frac{\theta}{2}) \sum_{k=0}^{N-1} v_k^2 +  S/2 \sum_{k=0}^{N-1} |v_k|,
\label{impactExprWithTemporaryTerm}
\end{equation}

If one assumes that all $v_k$ have the sign of $X$, the last term becomes equal to $X S/2$. The optimal solution, i.e. the one that minimizes the expected impact costs, is

\begin{equation}
\mathbf v^{*} \equiv \arg \min_{\mathbf v} E\left[C (\mathbf v)\right] = \left(\frac{X}{N},  \frac{X}{N}, ... ,  \frac{X}{N}\right)^T.  
\label{eq:solution_alm_chriss}
\end{equation}
showing that the solution simply consists in  trading at a constant rate over the periods. 
\\
The variance of the execution cost is 
\begin{equation}
 Var\left[C (\mathbf v)\right]  = E \left[ {\left(C (\mathbf v) -  E[C (\mathbf v)] \right)}^2\right] 
= E\left[ {\left( \sum_{k=0}^{N-1} \eta_k \sum_{j=k}^{N-1} v_j  \right)}^2 \right],
\label{varianceExprBrutta}
\end{equation} 
We assumed the $\eta_k$ are independent, so we have $E[\eta_i \eta_j] = 0$ for $i \neq j$ and thus:
\begin{equation}
  Var\left[C (\mathbf v)\right]  = \sigma^2 \sum_{k=0}^{N-1} (\sum_{j=k}^{N-1} v_j )^2.
  \label{varianceExpr}
\end{equation}

The second innovation in Almgren and Chriss is that they consider risk aversion for optimal execution: they minimize the sum of expected cost and costs' risk. By mimicking the theory on portfolio optimization \cite{markowitz1968portfolio,elton1995persistence}, Almgren and Chriss consider as optimal trading schedule the solution of
\begin{equation}
 \arg \min_{\mathbf v} \left(E[C (\mathbf v)] + \lambda Var[C (\mathbf v)] \right).
 \label{eq:lagrange_almgren_chriss}
\end{equation}
where $\lambda$ is the coefficient of risk aversion. The higher is the $\lambda$, the more important is risk with respect to cost. A risk neutral investor corresponds to $\lambda=0$. The set of solutions to this problem for different values of $\lambda$ is called \emph{optimal frontier}.

By using the impact model of Eqs. \ref{priceDyn} and \ref{eq:almgren_chriss_effective_price} one obtains
\begin{equation}
 v_k = A \cosh(\beta (T - t_k)), 
 \label{eq:optimal_solution_alm_chr}
\end{equation}
where $A$ is a normalization constant. 
The coefficient $\beta$ is related to the coefficient of risk-aversion $\lambda$, and is given by\footnote{The equation found in \cite{almgren2001optimal} is slightly different because they define the temporary impact and the variance as proportional to the interval length $\tau = T/N$.}:
\begin{equation}
\beta = \sqrt{\frac{\lambda \sigma^2}{\rho}} , \   \    \    \lambda > 0.
\end{equation}

The solution of Eq. \ref{eq:optimal_solution_alm_chr} shows that the more risk averse is the investor (i.e. the higher is $\lambda$), the more front loaded is the trading schedule. This means that more volume is traded at the beginning of the execution in order to minimize the uncertainty on execution price of the last part of the trading schedule.

\section{Data}

For our empirical research we have used two different datasets. The first one reports the trading activity at the London Stock Exchange (LSE) and covers from May 2000 to December 2002.
The database allows to reconstruct the order book and thus there is no ambiguity in the data, especially for identifying the initiator of each trade. We present here the data of two stocks, namely Astrazeneca (symbol AZN) and Vodafone (symbol VOD).

The second dataset contains a small sample of two NASDAQ stocks, namely Apple (AAPL) and Amazon (AMZN), for the period July-August 2009. The database contains trades and quotes and in this case we infer the initiator of the trades by using the Lee and Ready algorithm  \cite{lee1991inferring}. The algorithm consists in comparing the trade price with the prevailing quote and in identifying the trade as buyer (seller) initiated if the trade price is higher (lower) than the mid price of the prevailing quote. The case where trade price equals mid price remains undetermined, but it represents a very small fraction of the total (smaller than $ 0.1 \%$). 
Table \ref{tab:presentation_of_the_four_stocks} shows some summary information of the selected stocks.

\begin{table}[htbp]
   \centering
   \begin{tabular}{@{} lccccc@{}} 
      \toprule
      	&  Number & Avg. num. of &  Avg. price&  Avg. tick & Avg. tick size -\\
      Symbol     & of days & trans. in a day &   ($\pounds$ or \$)&  size  ($\pounds$ or \$)& price ratio (bp)\\
      \midrule
      AZN      & 675 & 731 & 26.65 & 0.01 & 3.7 \\
      VOD      &  675 & 1360 &  2.95 & 0.0025 &  8.4 \\
    \hline
      AAPL & 42 & 24094 & 157.06 &0.01 & 0.64 \\
      AMZN & 42 & 13890 & 83.87 & 0.01 & 1.2 \\
      \bottomrule
   \end{tabular}
   \caption{Summary information of the investigated stocks.}
   \label{tab:presentation_of_the_four_stocks}
\end{table}

\section{Propagator model in trade time}

In order to solve the problem of the optimal execution, it is necessary to choose a model for the price dynamics. The most delicate aspect is to model in a careful way how price reacts to our trades, i.e. we need to have a good model of market impact. Here we propose to adapt a market impact model recently introduced \cite{bouchaud2004fluctuations} and termed propagator model. This model takes into account both the volume and the temporal dependence of the impact. In fact, it has been shown that price return reacts to a trade not only simultaneously to the trade, but also later in the future, and this effect is slowly decaying. In the next section we will review the propagator impact model.

It is important to stress that the propagator model has been proposed and calibrated on transaction by transaction data. In the optimal execution scheme presented in the previous section, the length $\tau$ of the time intervals corresponds to several minutes or the corresponding average number of trades. Therefore we need to adapt the model to describe impact in macroscopic time and test it on the data. This will be done in the next Section.

 The propagator model for market impact describes the price dynamics in transaction time, i.e. time  $t_n$, $n \in \mathds{N}$ increases of one unit after each transaction. The price $p_n$ is the log mid-price of the asset right before the transaction at time $t_n$. Let $v_n$ be the signed volume of trade at time $t_n$.

The propagator model describes the price dynamics as 
\begin{equation}
 p_n = p_0 + \sum_{k = 0}^{n-1} \left[\mathcal{G}(j, k, v_{k}) + \eta_k\right]
 \label{eq:first_equation_propagator_ch_3}
\end{equation}
where $\mathcal{G}$ is a general impact function which describes the impact at time $t_n$ of the trade of volume  $v_k$ at time $t_k$ ,
$\eta_k$ is an independent and identically distributed random term with mean $0$ and variance $\sigma^2$ and $p_0$ is the log mid price at a time $t_0$ far in the past\footnote{In the original formulation this time was set at $-\infty$. Here we consider a generic initial time (often set as the beginning of the trading program).}. In order to ensure causality $\mathcal{G}(j, k, v_{k})$ is nonzero only for $j > k$. Moreover, Refs.  \cite{bouchaud2004fluctuations} showed from real data that the volume and time dependence of $\mathcal{G}$ can beapproximately factorised, i.e.
\begin{equation}
 \mathcal{G}(j, k , v_{k}) \simeq f( v_k) \cdot G_0(j, k).
\end{equation}
Finally,  we assume that $G_0$ is translationally invariant \cite{bouchaud2004fluctuations}, i.e.
\begin{equation}
G_0(j, k) \equiv G_0(j - k).
\end{equation}
Since $\mathcal{G}$ is non-zero only for $j > k$ we have that $G_0(k)$ is non-zero only for $k > 0$. 

Thus  the model of price dynamics is a geometric random walk with market impact, i.e.

\begin{equation}
 p_{j} = p_0 + \sum_{k = 0}^{j-1}[\eta_k +  f(v_k) G_0(j-k)],
 \label{tempImpactFinalFunction}
\end{equation}
The difference between two consecutive log prices (i.e. the log-return) is
\begin{equation}
 r_{j} \equiv p_{j+1} - p_j =  G_0(1)f(v_j) + \sum_{k>0}[G_0(k+1) - G_0(k)]f(v_{j-k})+ \eta_j.
 \label{consecutivePrices}
\end{equation}
The return $r_{j}$ can be divided into three components: the impact of the last transaction, which is proportional to  $f(v_j)$, the effect of the decay of the impacts of all older transactions $\propto f(v_{j-k})$, and a random shock $\eta_j$.

The propagator $G_0$ of real transaction by transaction data is well fitted by the function \cite{bouchaud2004fluctuations,bouchaud2006random}
   \begin{equation}
 G_0(l) = \frac{\Gamma_0}{(l_0^2 + l^2)^{\beta/2} },
 \label{BouchaudProposedG_0}
\end{equation}
where $\Gamma_0$ is a multiplicative constant, $l_0$ is a correction for small lags and $\beta$ the coefficient of the power law decay for $l \gg l_0$.

The factorization of $\mathcal{G}$ in a volume and a time component leads to the definition of the instantaneous impact function $f(v)$. This function is the tick by tick expected return due to a trade initiated by a given signed volume  $v$, i.e.
\begin{equation}
f(v) = E[r | v],
\label{eq:impact_function_def}
\end{equation}
Data indicates that $f(v)$ is well approximated by an odd function, i.e. $f(-v) \simeq -f(v)$. 

Moreover, many studies \cite{bouchaud2004fluctuations}, \cite{lillo2004long}, \cite{bouchaud2009markets} have shown that $f$ is a concave function of the volume $v$ and is typically well fitted by a logarithmic function or a power law function with exponent smaller than one. Note however that this is true for individual trades returns. As returns are computed on longer time scales, the return conditional to volume imbalance becomes more and more linear (see Ref. \cite{bouchaud2009markets}).

\section{The propagator model under other time measures}

The propagator model was originally introduced to describe the transaction by transaction dynamics of price. However, in order to use this model for the optimal execution problem as stated above, we need to have a model in aggregate time, i.e. where time advances by one unit after one interval $\tau$. First of all, notice that there are several possible definitions of aggregated time in financial data. The first, and most natural one, is obtained by considering $\tau$ as a fixed real time interval, for example five minutes. This choice brings sometime problems related to the fact that financial processes are often not homogeneous in real time, because of periodicities, trading clustering, etc. For this reason one can consider also other time measures, such as aggregated trade (or transaction) time and aggregated volume time. In the former case, time advances by one unit any time a certain fixed number of trades have been executed. In the second measure, time advances by one unit when a given volume has been traded in the market. In this section, we will show how to adapt the propagator model to real time.  In the Appendix we present the case of aggregated trade time.

\subsection{Calibration of the propagator model in real time}
\label{subsec:constant_time_intervals} 
We fix $t_0$ as the opening time of the market  and we define the series of times $t_n$ as separated by a constant time interval. For definiteness in the following we will consider 5 minute intervals, even if similar results apply to different durations. In this case $t_0 =$ 8:00, $t_1 =$ 8:05, $t_2 =$ 8:10, ... As above, $p_n$ is the log mid price right before time $t_n$.  We define the series of aggregated volumes $v_n$ in terms of the volumes $v^{tt}_i$ of the single transactions ($tt$ stands for transaction by transaction), i.e. 
\begin{equation}
 v_n =  \sum_{[t_n , t_{n+1}]} v^{tt}_i.
\end{equation}
by this notation we mean that we sum the all signed volumes, $v^{tt}_i$, that are traded between time $t_n$ and $t_{n+1}$.
The quantity $v_n$ is the volume imbalance over the $n$-th time interval. This quantity may be very big in absolute value during time intervals of high market activity, like, for example, near market opening and closing. Therefore, we consider a related variable, the \emph{normalized} volume imbalance $v^{nor}_n$, which is obtained by dividing the volume imbalance $v_n$ in the $n$-th interval by the total absolute volume traded in the interval,
\begin{equation}
 v^{nor}_n =  \frac{\sum_{[t_n , t_{n+1}]} v^{tt}_i}{\sum_{[t_n , t_{n+1}]} |v^{tt}_i|}
 \label{def:impact_volumes_interval_time}
\end{equation}
When there is no trade in the $n$-th interval, we define $v^{nor}_n =0$. Note that $v^{nor}_n \in [-1, 1]$. This definition has also been used in Ref. \cite{lillo2004long}.
We can finally state the propagator model in the new time setting:
\begin{equation}
 p_{j} = p_0 + \sum_{k = 0}^{j-1}[\eta_k +  f(v^{nor}_k) G_0(j-k)],  \;\;\;\; \eta \sim IID
 \label{tempImpactFinalFunctionRealTime}
\end{equation}
The difference between two consecutive log prices (log-return) is
\begin{equation}
 r_{j} \equiv p_{k+1} - p_k  = \sum_{k=0}^{j-1} g(k) f(v^{nor}_{j-k})+ \eta_j.
 \label{consecutivePricesRealTime}
\end{equation}
where we defined $g(k) \equiv G_0(k+1) - G_0(k)$, and $G_0(0) = 0$. 
 
\begin{figure}[htbp]
\begin{center}
\includegraphics[height=5.5cm]{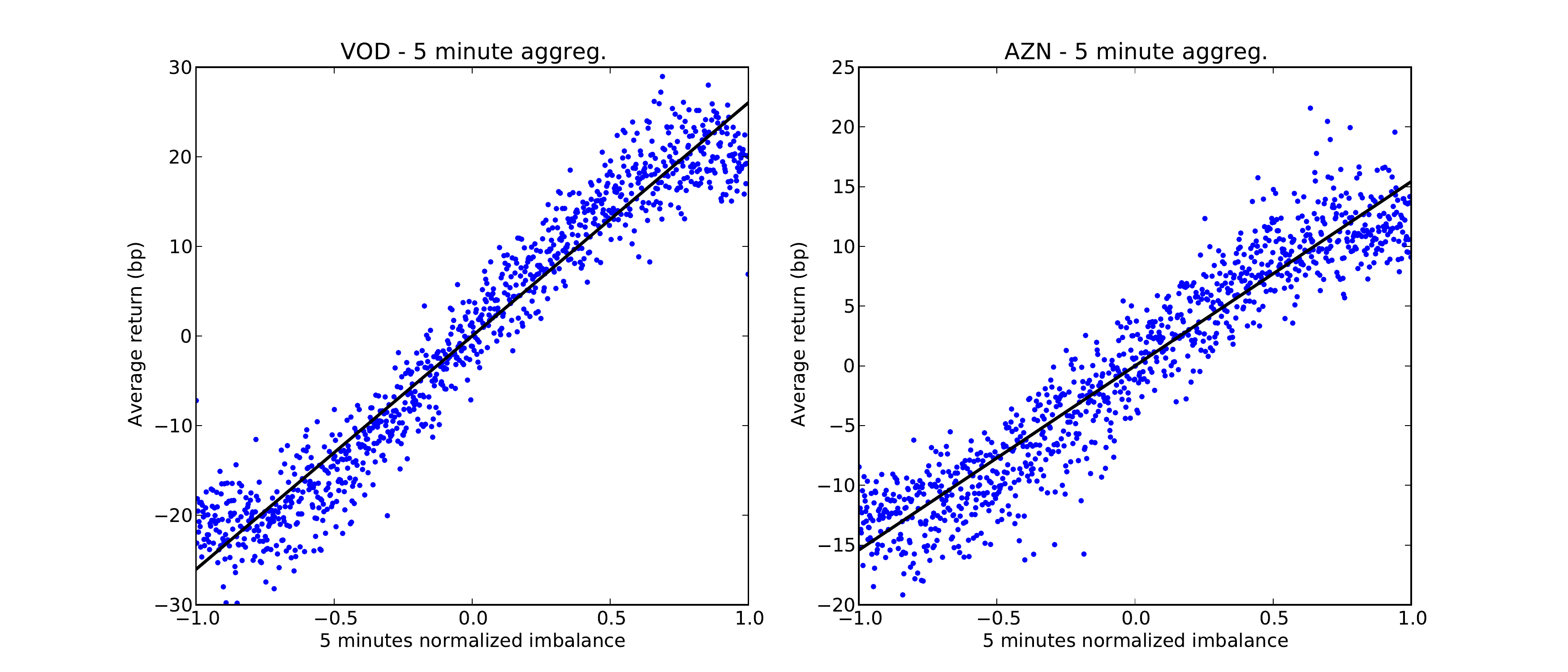}
\includegraphics[height=5.5cm]{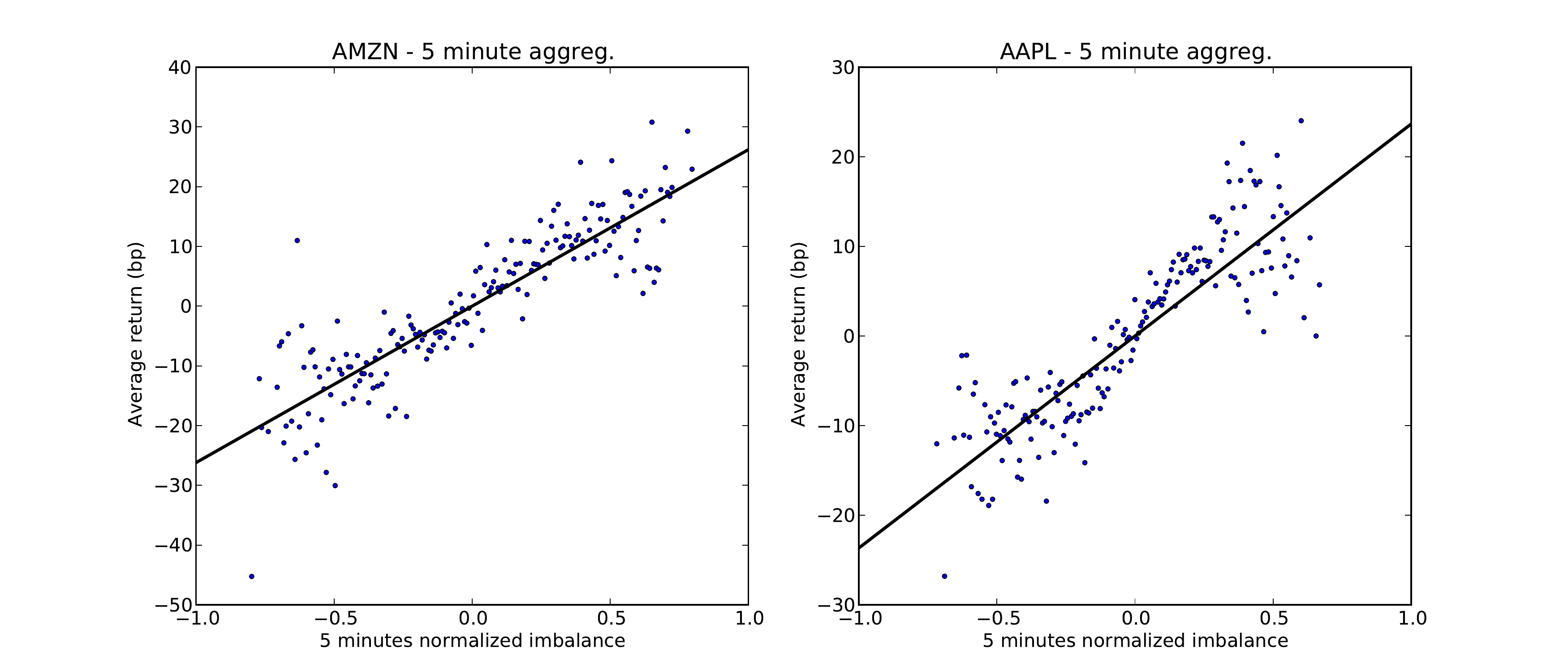}
\end{center}
\caption{Estimation of the normalized impact of Eq. \ref{fvnor} in real time. Parameters of the fits are given in Table \ref{tab:estimated_parms_real_time_G_0}..
}
\label{RealTime_5_mins_TimeAggreg_ImpactFitted}
\end{figure}

We now calibrate the propagator model on the four investigated stocks.
We first estimate the impact function $ f(v^{nor})$ of the normalized volume imbalance. It is defined as the expected return in a five minutes interval conditioned to a normalized volume imbalance $v^{nor}_n$ in the same interval, i.e.
\begin{equation}
 f(v^{nor}) = E [r_n | v^{nor}_n].
 \label{fvnor}
\end{equation}
In figure  \ref{RealTime_5_mins_TimeAggreg_ImpactFitted} we show the scatter plots of $ f(v^{nor})$. As it is immediately evident, on this time scale and having used the normalization procedure, the impact function becomes very close to a linear function of the normalized volume. This will be a key ingredient to find the optimal execution, because the linearity of the impact allows an analytical solution of the optimization problem. For this reason we fit the impact function in real time with a linear function given by
\begin{equation}
 f(v^{nor}) = \theta v^{nor}.
 \label{eq:impact_function_nor_lin}
\end{equation}
Table \ref{tab:estimated_parms_real_time_G_0} reports the results of the fit for the four stocks.

\begin{figure}[htbp]
\begin{center}
\includegraphics[height=5.5cm]{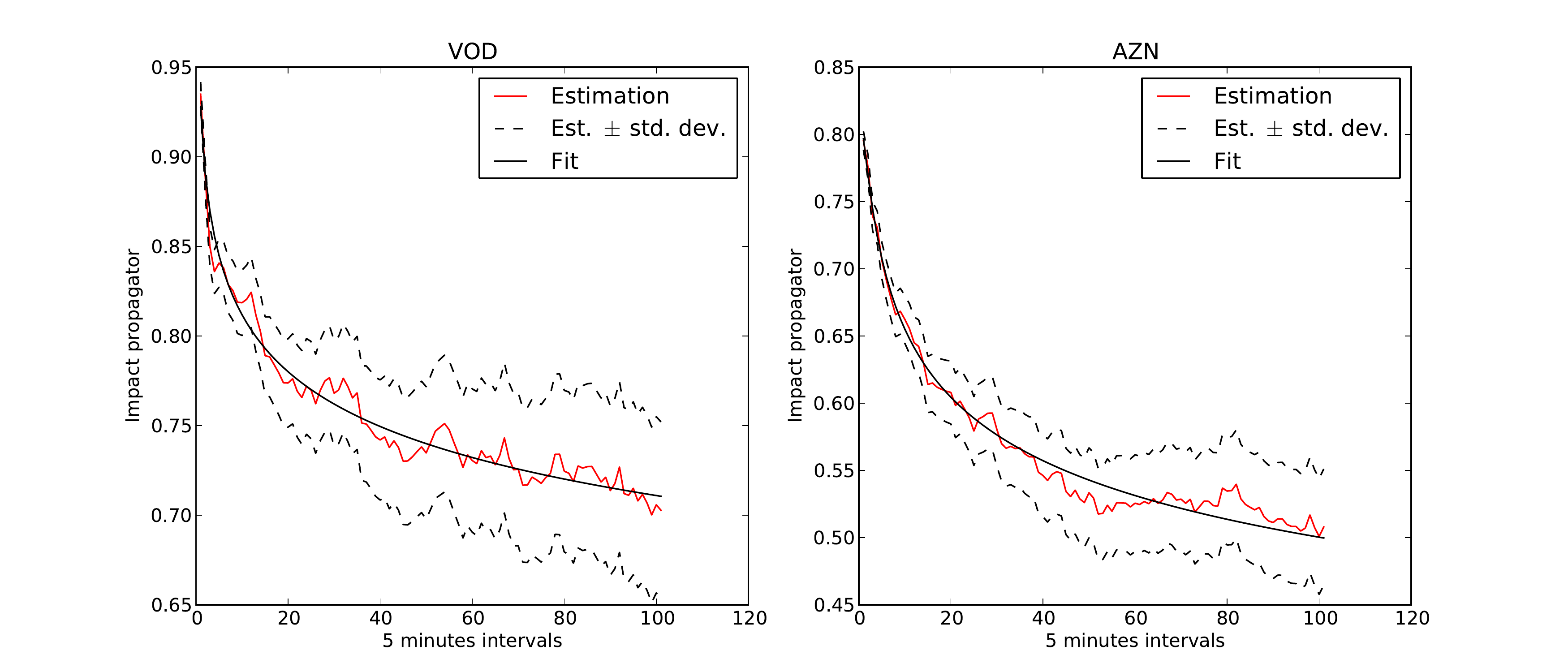}
\includegraphics[height=5.5cm]{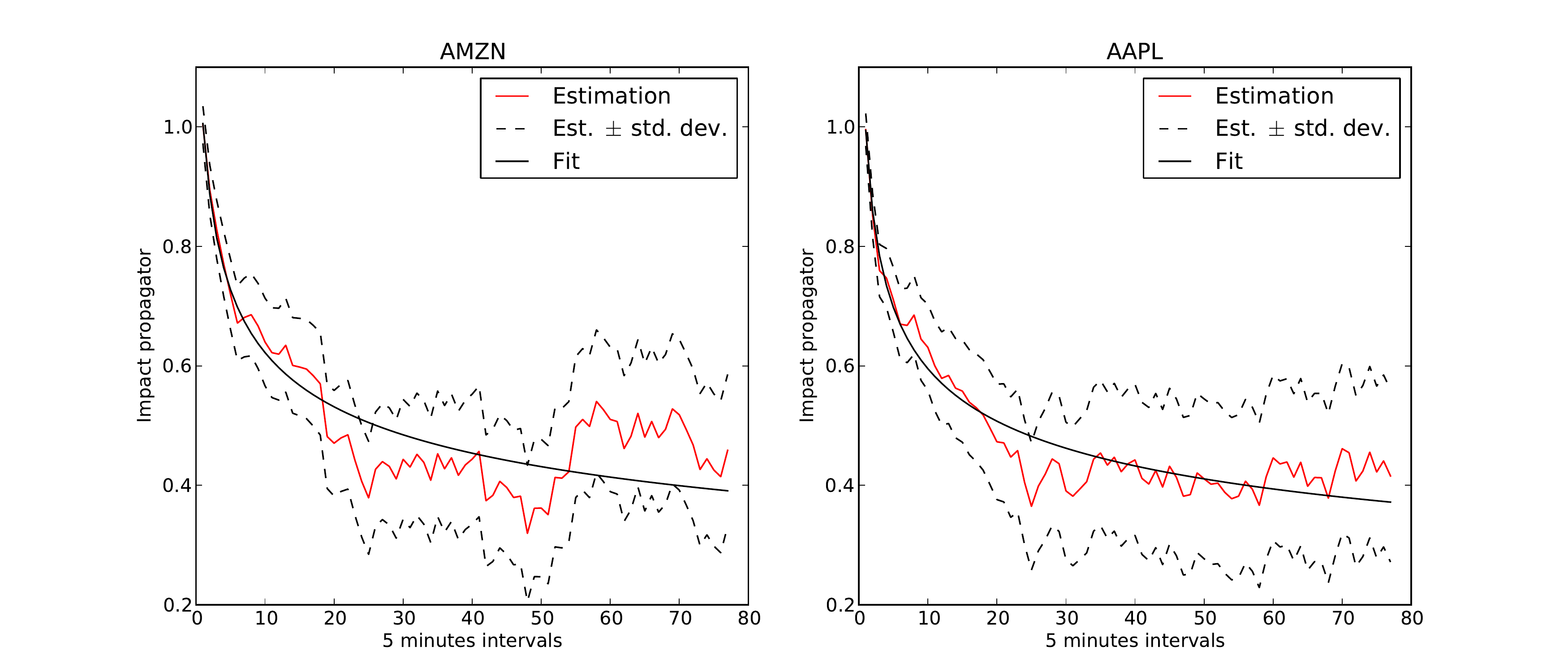}
\end{center}
\caption{Estimation of the impact propagator $G_0$ in real time five minute intervals. Each point is the result of binning many transactions with similar normalized imbalance and the y axis is the expected return. Parameters of the fits are given in Table \ref{tab:estimated_parms_real_time_G_0}.
}
\label{RealTime_5_mins_G_0_fitted}
\end{figure}

We can now estimate the impact propagator $G_0$ in real time, by using the linear function $f(v^{nor})$ obtained above. 
We use eq. \ref{consecutivePricesRealTime} as a linear regression model to find the best coefficients $g(k)$ and from those obtain $G_0$.
From figure \ref{RealTime_5_mins_G_0_fitted} we observe that for both datasets the propagator model seems to fit quite well the data. The fit is quite good also for the NASDAQ stocks, where the sample is small. This encourages to think that one does not need long historical datasets in order to fit the propagator model.

We perform a best fit of the propagator with the functional form of Eq. \ref{BouchaudProposedG_0}. 
The parameters of the best fits are reported in Table \ref{tab:estimated_parms_real_time_G_0}. Below we will use impact propagators with these parameters in order to calibrate the optimal trading schedules. 

\begin{table}[htbp]
   \centering
   \begin{tabular}{@{} lcccc @{}} 
      \toprule
      & & \multicolumn{3}{c}{$G_0$ parameters} \\
      \cmidrule(lr){3-5} 
      Symbol & $\theta$ (bp) & $\Gamma_0$  & $\l_0$ & $\beta$ \\
      \midrule
      AZN     & $15.4 \pm 0.5$ & $1.40 \pm 0.04 $ & $20 \pm 1$ & $0.190 \pm 0.006$ \\
      VOD      &  $26.0 \pm 0.8 $& $1.07 \pm 0.01 $& $ 4 \pm 1$& $0.075 \pm  0.002$ \\ 
      AAPL & $21.9 \pm  0.9 $& $1.01\pm 0.02$& $0.41\pm 0.27$& $0.23\pm  0.007$ \\
      AMZN & $26.9 \pm  0.8 $& $1.05 \pm 0.03$& $0.70 \pm 0.35$ & $0.23 \pm 0.011$ \\ 
      \bottomrule
   \end{tabular}
   \caption{Parameters of the fits of the market impact propagator model in the real time setting, which we use in the impact model for the optimal execution. The parameters  are those obtained from the fits shown in figure \ref{RealTime_5_mins_TimeAggreg_ImpactFitted} and \ref{RealTime_5_mins_G_0_fitted}.}
    \label{tab:estimated_parms_real_time_G_0}
\end{table}

\subsection{Variance explained by the model}
\label{sec:variance_explained}
We conclude this section by discussing more quantitatively how well the propagator model fits the data for different time definitions and time intervals. In particular, for each regression we compute the $R^2 \in [0,1]$ coefficient, which is the fraction of the total variance  explained by the model. We summarize in Table \ref{tab:R_squared_coefficients} the $R^2$ coefficient for all the studied models. 
Apart from the cases discussed above, we include also the propagator fitted in trade time (as in the original paper of Bouchaud et al.), for aggregated transaction time with $d=8$ and $d=64$ transactions, and for real time with $1$ minute intervals. 

\begin{table}[htbp]
   \centering
   \begin{tabular}{@{} lccccc @{}} 
      \toprule
       & \multicolumn{5}{c}{$R^2$ coefficients of the linear regression} \\
      \cmidrule{2-6} 
      Symbol & trade  & 8 aggregated  & 64 aggregated & 1 minute & 5 minutes  \\
       & time& transactions  & transactions & aggregation & aggregation  \\
      \midrule
      AZN     &  0.224 & 0.245 & 0.304 & 0.192 & 0.204\\
      VOD      & 0.257 & 0.329 & 0.440 & 0.209 & 0.292\\ 
       AMZN&	0.172&	0.253&	0.229&	0.163&	0.153\\
      AAPL&	0.202&	0.302&	0.319&	0.223&	0.212\\
       \bottomrule
   \end{tabular}
   \caption{$R^2$ coefficients of the regression of the propagator model for different time definitions.}
    \label{tab:R_squared_coefficients}
\end{table}

We note that $R^2$ increases by increasing the length of the time interval (in number of transactions or in real time). As an extreme case note that for VOD in the 64 aggregated transactions model, $44 \%$ of the total variance is explained by the model. Also the NASDAQ stocks are quite well fit by the model, even if when aggregation increases, the quality of the fit remains roughly constant, probably because of the smallness of the sample.

\section{Optimal execution model}

We now present the theoretical solution of the optimal execution in the presence of transient impact. For the case where bid ask spread costs are neglected we re-obtain the results presented recently in Ref. \cite{bouchaud2009markets,alfonsi2009order}. We define the logarithmic transaction costs $c(\mathbf v)$, where $\tilde{p}_k$ is the log of the effective price $\tilde{P_k}$

\begin{equation}
c(\mathbf v) \equiv 
  \sum_{k=0}^{N-1} v_k (\tilde{p_k} - p_0).
  \label{eq:log_trans_cost}
\end{equation}

We note that in the limit of short time horizons of the execution $T$, the effective price $\tilde{P}_k$ does not deviate much from the starting value $P_0$. 
We can thus approximate:

\begin{equation}
 c(\mathbf v) =  \sum_{k=0}^{N-1} v_k \log\left(\frac{\tilde{P}_k}{P_0}\right)
     \simeq  \sum_{k=0}^{N-1} v_k \left(\frac{\tilde{P}_k - P_0}{P_0}\right)
     =  \frac{C(\mathbf v)}{P_0}.
\end{equation}
The logarithmic execution cost $c(\mathbf v)$ can therefore be thought as a fractional execution cost.
More important, minimizing the  execution cost $C(\mathbf v)$  is equivalent to minimizing the logarithmic execution cost $c(\mathbf v)$. 

\subsection{Baseline case: risk neutral investor without spread costs}

We are now ready to express the execution costs within the framework of the market impact propagator model described above.
We start from equation \ref{tempImpactFinalFunctionRealTime}, which we report here (dropping the superscript $nor$ of $v_k^{nor}$):
\begin{equation}
  p_{n} = p_0 + \sum_{k = 0}^{n-1}[\eta_k +  f(v_k) G_0(n-k)].
  \label{tempImpactFinalFunction_chap5}
\end{equation}

We  recall that the effective log-midprice $\tilde{p}_k$ is the logarithm of the average mid-price at which we assume to trade the shares $v_k$ between time $t_k$ and time $t_{k+1}$. The log price $p$ of the stock is a continuous time process, which we sample every 5 minutes to obtain the discrete price series $p_n$. Therefore, it is reasonable to assume that the effective log midprice $\tilde{p}_k$ of the trade between time $t_k$ and $t_{k+1}$ is given by the average of the prices at the two times:
\begin{equation}
\tilde{p}_{k} = \frac{p_k + p_{k+1}}{2}.
\label{eq:effective_price_as_average}
\end{equation}
The equation that describes the dynamics of effective price is therefore 
\begin{equation}
  \tilde{p}_{n} = p_0 + \sum_{k = 0}^{n}[\eta_k +  f(v_k) \tilde{G}_0(n-k)] 
 \end{equation}
where we defined the effective propagator $\tilde{G}_0$ as
\begin{equation}
 \tilde{G}_0(0) = \frac{G_0(1)}{2},\   \    \tilde{G}_0(1) = \frac{G_0(1) + G_0(2) }{2},\   \    \tilde{G}_0(2) = \frac{G_0(2) + G_0(3) }{2},\   \  ... 
 \label{eq:definition_effective_prop}
\end{equation}
With this impact model the fractional execution costs $c(\mathbf v)$ is given by

\begin{equation}
c(\mathbf v)= 
\sum_{n=0}^{N-1} v_n \left[\sum_{k = 0}^{n} \left(\eta_k + f(v_k)\tilde{G}_0(n-k)\right)\right].
\label{eq:fraction_transaction_costs_basic_model}
\end{equation}
and its expected value is
\begin{equation}
 E[c(\mathbf v)]= \sum_{n=0}^{N-1} v_n [\sum_{k = 0}^{n}f(v_k)\tilde{G}_0(n-k)]
 \label{eq:expect_value_fraction_transaction_costs_basic_model}
\end{equation}

Let us assume for simplicity that  that $f$ is linear, even if it can be time dependent, i.e.
\begin{equation}
 f(v_k) = \theta_k v_k.
 \label{eq:impact_func_varied_coefficients}
\end{equation}
where $\theta_k$ are coefficients describing a possible deterministic change of impact during the trading day.  
As we have seen above, for five minute intervals the linearity of $f$ is a reasonable approximation.

We now define the $N \times N$ triangular impact matrix as
\begin{equation}
  \mathcal{I}_{i,j} \equiv  \left\{
     \begin{array}{lr}
       \theta_i \tilde{G}_0(i-j) & i \geq j  \\
       0 & i < j
     \end{array}
   \right.
   \label{eq:impact_matrix_definition}
\end{equation}
which allows us to rewrite the expected logarithmic execution costs as
 
\begin{equation}
  E[c(\mathbf v)]= \mathbf v^T \mathcal{I}  \mathbf v. 
  \label{eq:impact_costs_basic}
\end{equation}

We want to find the optimal trading schedule, i.e. the one that minimizes

\begin{equation}
 \min_{\mathbf v} \mathbf v^T \mathcal{I}  \mathbf v,
\end{equation}
subject to the constraint
\begin{equation}
 \sum_k v_k = \mathbf 1^T \mathbf v =  X.
\end{equation}
where $\mathbf 1$ is a vector whose $N$ elements are all 1. This a quadratic minimization problem, analogous to the well known Markowitz best portfolio problem (see \cite{markowitz1968portfolio,elton1995persistence}). We use a Lagrange multiplier $z$ to enforce the constraint and we obtain the following unconstrained minimization problem

\begin{equation}
 \min_{\mathbf v} \Lambda (\mathbf v, z) =  \min_{\mathbf v} (\mathbf v^T \mathcal{I}  \mathbf v + z \cdot ( \mathbf 1^T \mathbf v - X) ).
\end{equation}
The optimal solution $\mathbf v^\star$ is 
\begin{equation}
 \mathbf v^\star =  - \frac{z}{2}\   \mathcal{I} ^{-1} \mathbf 1 =  \frac{X }{\mathbf{1}^T \mathcal{I}^{-1} \mathbf{1}} \mathcal{I}^{-1} \mathbf{1},
 \label{eq:markowitz_solution_first_step}
\end{equation}
and its expected execution costs:

\begin{equation}
  E[c (\mathbf v^\star)]= \mathbf v^{\star T} \mathcal{I}  \mathbf v^\star = 
  X^2 \frac{( \mathbf{1}^T \mathcal{I}^{T-1}) \mathcal{I} (\mathcal{I}^{-1} \mathbf{1})} {(\mathbf{1}^T \mathcal{I}^{-1} \mathbf{1})^2} =
  \frac{X^2}{\mathbf{1}^T \mathcal{I}^{-1} \mathbf{1}}.
\end{equation}
We note that the execution costs are quadratic in the total volume $X$. Interestingly the optimal trading trajectory $\mathbf v^\star$ can be computed with a simple closed-form matrix expression.

\subsection{Optimal execution with spread costs}
\label{sec:bid-ask_spread_costs}
One important extension to the model of last section is to consider the contribution of bid ask spread costs. We assume that the bid-ask spread has a constant value\footnote{The model can easily extended to the case of a deterministically varying (average) spread. This extension could for example capture the intraday variation of spread.} $s = A - B$, where $A$ is the ask and $B$ the bid.
Recalling that $P = \frac{A + B}{2}$ is the stock mid price, we define $\delta$ as half of the fractional bid-ask spread, i.e.

\begin{equation}
\delta \equiv \frac{s/2}{P} = \frac{A - B}{A + B}.
\label{eq:delta_definition}
\end{equation}

Then we express the effective log-price $\tilde{p}_k$ is

\begin{equation}
 \tilde{p}_n = p_0 + \sum_{k = 0}^{n}[\eta_k +  f(v_k) \tilde{G}_0(n-k)] +\mbox{sign}(v_k)  \delta,
 \end{equation}
because we pay half the bid-ask spread on execution.

The expected value of the transaction costs including the bid-ask spread contribution is therefore
\begin{equation}
   E[c(\mathbf v)]= \mathbf v^T \mathcal{I}  \mathbf v + \delta \mathbf 1^T |\mathbf v|.
     \label{eq:impact_costs_plus_fixes}
\end{equation}
Minimizing Eq. \ref{eq:impact_costs_plus_fixes} is more complicated than minimizing Eq. \ref{eq:impact_costs_basic} because of the term with the absolute value of $\mathbf v$. 
We can not use anymore the machinery of Lagrange multipliers and matrix inverse and we will use numerical minimization methods.

\subsection{Optimal execution with risk aversion}
\label{sec:risk_aversion_theory}

In this section we consider an extension of the optimal execution model taking into account risk-aversion. 
The idea that risk matters in determining the optimal execution schedule was first proposed by Almgren \& Chriss in \cite{almgren2001optimal}. We express the variance $Var[c(\mathbf v) ]$ of the fractional transaction costs $c(\mathbf v)$ of equation \ref{eq:fraction_transaction_costs_basic_model}. Recalling that the $\eta_n$ are independent and identical distributed random variables with zero mean and variance $\sigma^2$, we have 
\begin{multline}
 Var[ c(\mathbf v)] 
 =
  E \left[ {\left(c( \mathbf v) - E[ c( \mathbf v)] \right)}^2\right] 
 =
 E[( \sum_{k=0}^{N-1} v_k  \sum_{j = 0}^{k-1}\eta_j)^2] 
 = \\
 = E[( \sum_{k=0}^{N-1} \eta_k  \sum_{j = k}^{N-1}v_j)^2] 
 =
   \sigma ^2 \sum_{k=0}^{N-1} (\sum_{j = k}^{N-1}v_j)^2.
 \label{sec:add_risk_averse-equa_var}
\end{multline}

The variance is bilinear in the trading schedule $\mathbf v$. We can therefore express it as
\begin{equation}
Var[c(\mathbf v)] = \sum_{k,j} \mathcal{V}_{k,j} v_k v_j,
\end{equation}
where $\mathcal{V}$ is
\begin{equation}
\mathcal{V} = \sigma^2 \cdot \begin{pmatrix} 
0 & 0 & 0 & \cdots & 0 \\
0& 1 & 1 & \cdots & 1 \\
0& 1 &  2 &\cdots & 2 \\
\vdots & \vdots & \vdots &\ddots & \vdots \\
0 & 1 & 2 & \cdots & (N-1)
 \end{pmatrix}.
 \label{eq:variance_matrix_definition}
\end{equation}

We now  formulate the problem of minimizing execution costs with a risk-aversion term by following  the construction of \cite{almgren2001optimal}. Specifically, we add the variance term multiplied by a coefficient $\lambda$ (which represents the risk aversion parameter) to the transaction costs of equation \ref{eq:impact_costs_plus_fixes}.
We obtain a total costs  function $ c_{\lambda}(\mathbf v)$
\begin{equation}
  c_{\lambda}(\mathbf v) \equiv E[c( v_t)] + \lambda Var [ c( v_t)] + \delta (\mathbf v) = 
 \mathbf v^T [\mathcal{I}+ \lambda \mathcal{V}] \mathbf v + \delta  \mathbf 1^T |\mathbf v| .
 \label{objFunctDiscrete}
\end{equation}
The minimum of the objective function in equation \ref{objFunctDiscrete} can be found using numerical  methods.

Note that if spread costs are negligible we can solve analytically the optimization problem and find the solution that minimizes the objective function \ref{objFunctDiscrete}. Setting  $\delta= 0$, we define $\mathcal{F} \equiv \mathcal{I}+ \lambda \mathcal{V}$.  By using again Lagrange multipliers, we have therefore the optimal trading schedule
\begin{equation}
\mathbf v^\star = z\   \mathcal{F} ^{-1} \mathbf 1 = \frac{ V}{\mathbf{1}^T \mathcal{F}^{-1} \mathbf{1}} \mathcal{F}^{-1} \mathbf{1}.
\end{equation}
 
\section{Empirically calibrated optimal execution strategies}

In this section we show a calibration to real data of the optimal execution schedules shown above. We will consider real time aggregation over five minute intervals.

In section \ref{subsec:constant_time_intervals} we showed how to fit the propagator model when using real time intervals. Here we point out that some care should be taken when defining the impact function to be used in the optimization. In fact, the impact function
\begin{equation}
 f(v_n^{nor}) = E [r_n | v^{nor}_n] =  \theta\   v_n^{nor}.
\end{equation}
gives the relation between price change and normalized volume imbalance.
We assume that the volume traded in the market has on average sign $0$, i.e. trades are equally likely buyer and seller initiated.  We define the total volume $W_k$ traded  between time $t_k$ and time $t_{k+1}$ as
\begin{equation}
 W_k = \sum_{ [t_k , t_{k+1}]} |v^{tt}_i|
\end{equation}
where $v^{tt}_i$ refer to the volumes of the individual transactions. This quantity equals the denominator of the equation 
\ref{def:impact_volumes_interval_time}. The volume we trade at step $k$ of the trading schedule gives a normalized imbalance of $ v^{nor}_k = {v_k}/{W_k}$ (because all the other volume has on average sign $0$). Therefore the impact function $f$ of the volume $v_k$ is
\begin{equation}
f(v^{nor}_k) =\frac{\theta}{W_k} v_k \equiv \theta_k v_k.
 \label{eq:impact_function_chap_8}
\end{equation}
which defines the series $\theta_k$. 
In the following we will approximate the series $W_k$ with a constant value, $W_k \equiv W$. In principle one could relax this assumption and take into account the intraday pattern of volume.
The parameter $\theta$ was estimated empirically and we report its value in Table \ref{tab:estimated_parms_real_time_G_0}.

In the following we will consider an investor that wants to execute a buy program in a trading day. We assume that the investor wants to execute a given fraction of the daily volume, and for definiteness we set this fraction at $1\%$. 
We define the participation rate $ \mathbf x$  as the $N$-dimensional vector, whose component $x_k$ is 
\begin{equation}
x_k \equiv \frac{v_k}{W_k} .
\label{eq:participation_rate_def}
\end{equation}
i.e. the ratio between the traded volume $v_k$ and the total market volume traded in the same interval, $W_k$. Given the $1\%$ volume target and the approximation $W_k\simeq W$, we have 
\begin{equation}
 X=\sum_{k=0}^{N-1} v_k = \frac{1}{100}  \sum_{k=0}^{N-1} W_k = \frac{N\cdot W}{100} ,
\end{equation}
Hence the participation rate $\mathbf x$ satisfies
\begin{equation}
 \sum_{k=0}^{N-1} x_k = \frac{N}{100} \Longrightarrow \langle x_k \rangle = 1 \%
\end{equation}

We use the functional form of Eq. \ref{BouchaudProposedG_0} for the impact propagator $G_0$. 

\subsection{Optimal execution without spread costs}
\label{sec:Solution_without_fixed_costs}

\begin{figure}[htbp]
\begin{center}
\includegraphics[height=5.5cm]{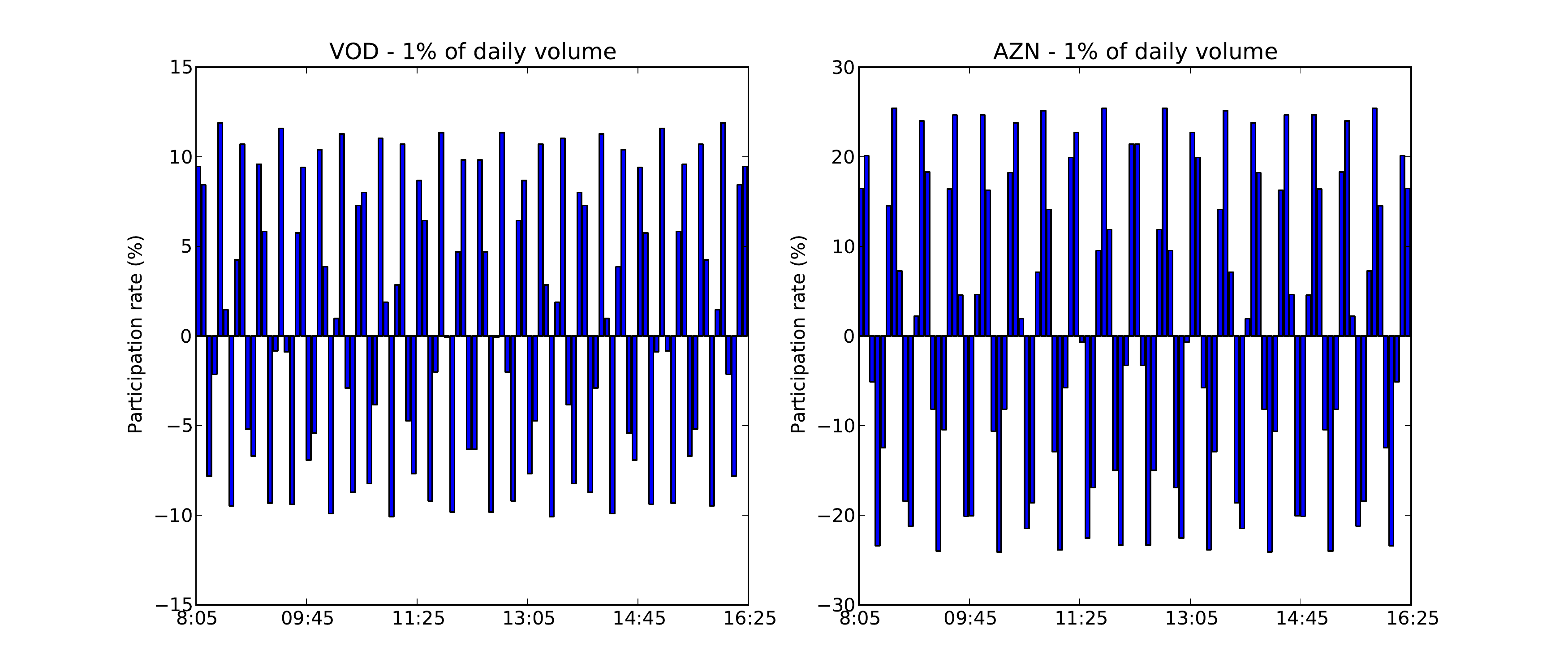}
\includegraphics[height=5.5cm]{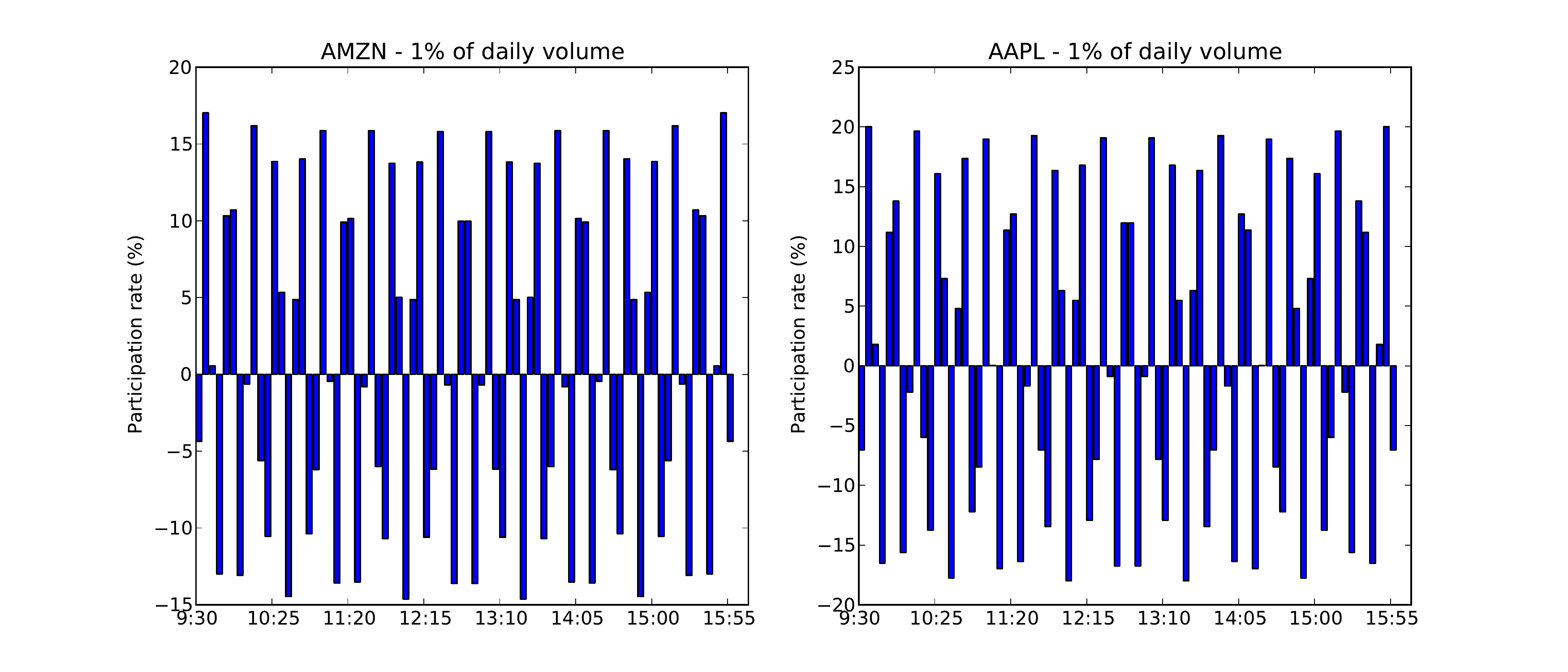}
\end{center}
\caption{Optimal execution strategies to buy 1\% of daily volume, neglecting the contribution of bid-ask spread and without risk aversion.
We plot the optimal participation rate over each 5 minutes interval.
}
\label{EmpiricalCalibration_OscillantSolutions_OptimalSolutionFlatLinear}
\end{figure}

We start with the simplest specification of the optimization problem by neglecting bid-ask spread costs and risk aversion. We computed analytically the optimal solutions $\mathbf v^\star$ for the four stocks. In figure \ref{EmpiricalCalibration_OscillantSolutions_OptimalSolutionFlatLinear} we plot the participation rate  $\mathbf x^\star$.

We see that the participation rate of figure \ref{EmpiricalCalibration_OscillantSolutions_OptimalSolutionFlatLinear} oscillates between positive and negative values, but its average value is indeed 1\%. The optimal solution consists in alternately buying and selling shares.  When an investor buys  he drives the price up.  Thus, if he sells right away he takes advantage of the higher price and moves it back down.\footnote{
We note that the total impact costs are in any case positive, so that one does not profit from market impact, as studied  in Ref. \cite{gatheral2009no}.}

However, the solutions of figure \ref{EmpiricalCalibration_OscillantSolutions_OptimalSolutionFlatLinear} are of little practical interest. 
In reality one does not trade at the mid-price of the stock, but buys at the ask price and sells at the bid (because we are considering execution with market orders). By buying, one therefore pays the mid-price \emph{plus} half the bid-ask spread.
By selling, one gets the mid-price \emph{minus} half the bid-ask spread. Thus, alternately buying and selling bears substantial costs of  bid-ask spread. We show in the next section how including the bid-ask spread costs changes dramatically the solution.

\subsection{Optimal execution with spread costs}
\label{sec:constant_bid-ask_spread}

We now solve the problem of optimal execution by including the cost of bid-ask spread.  We approximate the bid ask spread with a constant value $2\delta$ throughout the day, neglecting for simplicity the intraday pattern of spread. The expected value of the fractional transaction costs is 
\begin{equation}
   E[c(\mathbf v)]= \mathbf v^T \mathcal{I}  \mathbf v + \delta \mathbf 1^T |\mathbf v|,
     \label{eq:impact_costs_plus_fixes_chap8}
\end{equation}
where $\mathbf 1$ is the vector whose elements are all $1$.
The impact matrix $\mathcal I$ is the same as in section \ref{sec:Solution_without_fixed_costs} and we estimated the parameter $\delta$ from the data:\begin{equation}
 \delta_{AZN}=  5.27\  \mbox{bp},\   \   \   \delta_{VOD} = 10.12\   \mbox{bp}, \   \   \   \delta_{AMZN} = 1.47 \mbox{bp}, \  \  \
\delta_{AAPL} = 0.52 bp
\end{equation}

\begin{figure}[htbp]
\begin{center}
\includegraphics[height=5.5cm]{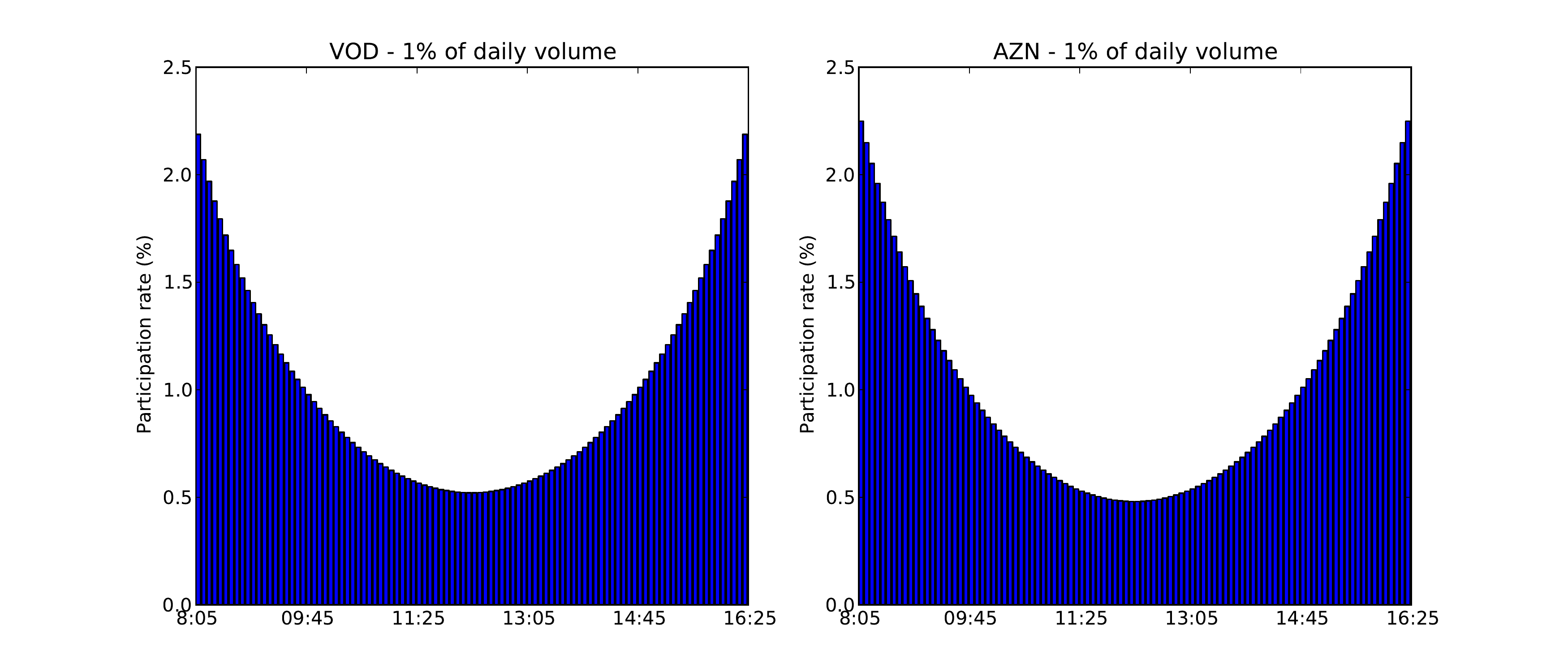}
\includegraphics[height=5.5cm]{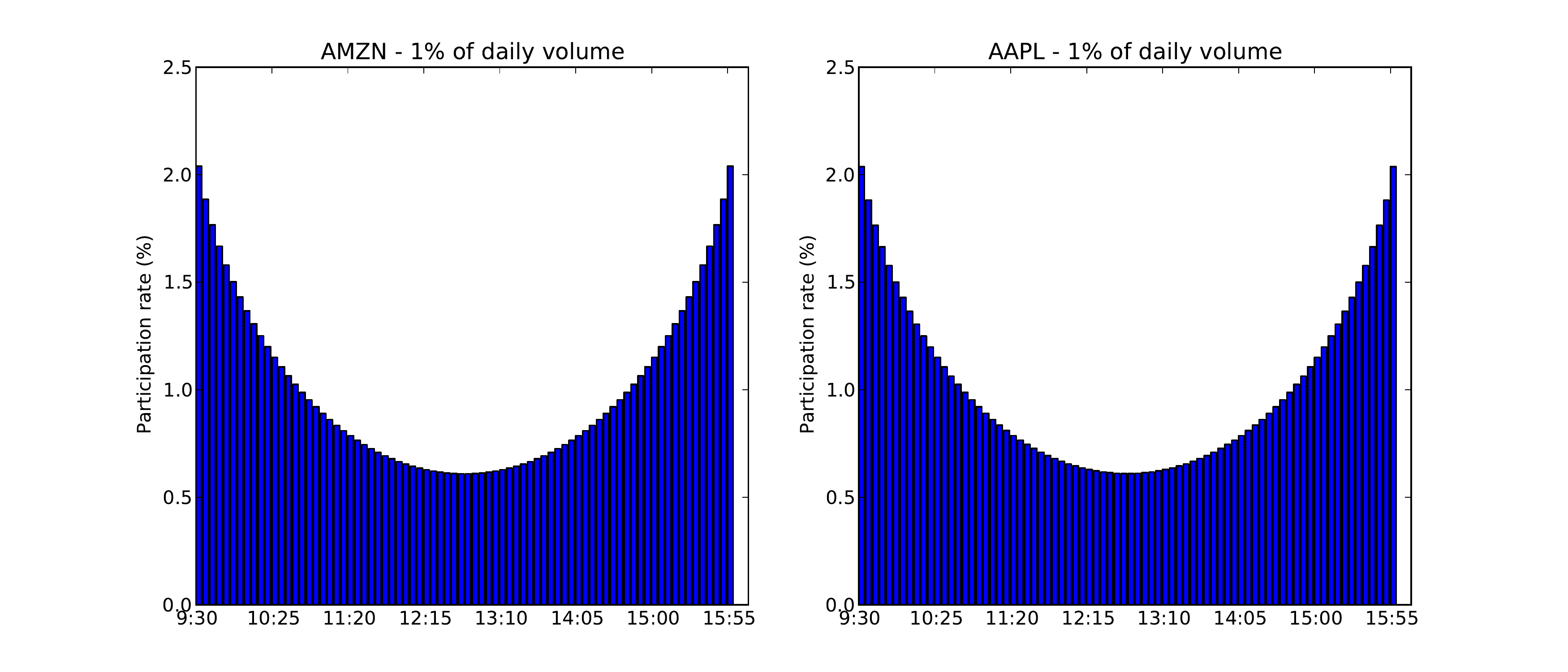}
\end{center}
\caption{Optimal execution strategies to buy 1\% of daily volume, when considering the contribution of bid-ask spread, but without risk aversion term.
We plot the optimal participation rate over each 5 minutes interval.
}
\label{EmpiricalCalibration_ConstantSpread_OptimalSolutionFlat}
\end{figure}

As we noted in section \ref{sec:bid-ask_spread_costs}, we cannot express in a closed form the solution $\mathbf v^\star$ that minimizes Eq. \ref{eq:impact_costs_plus_fixes}. We instead applied a numerical optimization method, using the algorithm \emph{L-BFGS-B} (see \cite{byrd1995limited}, \cite{zhu1997algorithm}) from the optimization routines of \emph{SciPy} \cite{Scipy}. The resulting optimal participation rates $\mathbf x^\star$ are shown in Figure \ref{EmpiricalCalibration_ConstantSpread_OptimalSolutionFlat}.

We see that including the bid-ask spread contribution in the execution costs has radically changed the form of the optimal solution.  The participation rate $\mathbf x^\star$ (and thus the trading schedule) is positive over all the market day.  In other words, if one has to pay a bid-ask spread cost on every transaction it is not convenient to sell and buy back shares. We can see that the optimal solutions are \emph{U-shaped}, i.e participation rate is high near market opening and closing,  and low during the central hours of the day. This happens because the market impact propagator $G_0$ decays in time. In fact, the high participation rate at the opening impacts much the price.  Then the period of low activity lets the market absorb the impact and recover a more favorable price.    
The high participation rate at the end instead impacts the price in the future, outside the time horizon $T$ of the execution.

The regularization achieved by the bid ask term is similar to what happens in portfolio optimization. It is known that adding to Markowitz objective function a penalty proportional  to the sum of the absolute values of the portfolio weights stabilizes the solution and corresponds to an exclusion of short positions \cite{L1pnas,caccioli11}. Similarly here, the bid ask spread term (which is motivated by the type of considered executions) stabilizes the solutions of the optimal execution and excludes oscillating solutions. We have tested that by choosing a $\delta$ parameter much smaller than the fractional spread, one recovers the U-shaped solution.

We  compare the optimal solution $\mathbf v^\star$ just obtained with the flat solution of Bertsimas and Lo \cite{bertsimas1998optimal}. It consists of constant $v_k=X/N$ over  each interval. In Table \ref{tab:summary_table_chap_8} we show a summary of the execution costs of the optimal solutions analyzed, in particular the fractional (per share) spread costs  $\bar{c}_{sp}(\mathbf v)$ and the fractional  impact costs   $\bar{c}_{imp}(\mathbf v)$, defined as
\begin{equation}
 \bar{c}_{sp}(\mathbf v) = \frac{ \delta \mathbf 1^T |\mathbf v|}{| \mathbf 1^T \mathbf v|},~~~~~~~~~~~~~ 
 \bar{c}_{imp}(\mathbf v) =  \frac{\mathbf v^T \mathcal I \mathbf v}{| \mathbf 1^T \mathbf v|}.
 \label{barc}
\end{equation}

We show these costs for all four stocks and three different solutions, namely  the \emph{flat} solution Eq. \ref{eq:solution_alm_chriss} of \cite{bertsimas1998optimal,almgren2001optimal}, the \emph{U-shaped} solution and the \emph{oscillating} solution of section \ref{sec:Solution_without_fixed_costs}.

\begin{table}[htbp]
   \centering
   \begin{tabular}{@{} lcccccc @{}} 
      \toprule
      \toprule
      & \multicolumn{6}{c}{Execution costs  per share, 1\% average daily volume} \\
      \cmidrule(lr){2-7}
       & \multicolumn{3}{c}{Fractional impact costs (bp)}  & \multicolumn{3}{c}{Fractional spread costs (bp)} \\
      \cmidrule(lr){2-4}
       \cmidrule(lr){5-7} 
 & flat & U-shaped & oscillating & flat & U-shaped & oscillating \\
      Symbol  & solution  & solution  & solution & solution  & solution  & solution  \\
      \midrule
      AZN      & 4.36 &  4.29  & 4.20 & 5.27 & 5.27 & 73.52 \\
      VOD       & 9.82 & 9.76 & 9.67 &10.12  & 10.12  & 81.62\\
      AMZN   & 4.09        & 4.03     & 4.02      & 1.47              & 1.47         & 14.20\\
	AAPL     & 3.17 &     3.12     & 3.12      & 0.52                  & 0.52      & 6.15\\
      \bottomrule
      \bottomrule
   \end{tabular}
   \caption{ 
  Comparison of  the fractional cost of impact and  fractional cost of spread (see Eq. \ref{barc}) for three solutions of the optimization problem, namely  the flat solution of Bertsimas and Lo \cite{bertsimas1998optimal}, the U-shaped solution and the oscillating solution.}
   \label{tab:summary_table_chap_8}
\end{table}

We can see that if we are concerned with the minimization of the impact costs alone, the oscillating solutions are indeed the best ones. 
However, when we consider the contribution of bid-ask spread, the oscillating solutions become much less attractive: by selling shares and buying them back one incurs in very high spread costs. Instead, the flat and U-shaped solutions have fractional spread costs per share equal to $\delta$ (because all elements of $\mathbf v$ have the same sign). The comparison between the  U-shaped solution and the flat one shows that the former has impact costs between 1 and 2\% lower.

\subsection{Risk aversion and efficient frontier of optimal execution}
\label{sec:empirical_efficient_frontier}

We now consider optimal execution by including a risk aversion term in the objective function. The objective function is (see Eq. \ref{objFunctDiscrete})
\begin{equation}
 c_{tot}(\mathbf v) = 
 \mathbf v^T [\mathcal{I}+ \lambda \mathcal{V}] \mathbf v + \delta  \mathbf 1^T |\mathbf v|.
 \label{objFunctDiscrete_chap8}
\end{equation}
To express the variance matrix $\mathcal V$ we need to estimate the variance $\sigma^2$ of the random shocks $\eta_n$. 
We calibrate it empirically as:
\begin{equation}
\bar{ \sigma}^2 = \left(\sum_{n = 0}^{L} \eta_n^2 \right)/{L},
\end{equation}
where $L$ is the total number of intervals in the dataset.
The resulting average variances $\bar{\sigma}^2$ for the four stocks are\footnote{
We note that the variance $\sigma^2$ has an intraday profile: there are intervals (e.g. the ones near the closing time) whose returns have statistically higher variance than the returns over other intervals. We will not model this effect.
}:
$$
 \bar{\sigma}^2_{AZN}=  350.81\  (\mbox{bp})^2,\   \   \   \bar{\sigma}^2_{VOD} = 764.52\  (\mbox{bp})^2,
 $$
 $$
\bar{\sigma}^2_{AMZN} = 395.62 \  {(\mbox{bp})}^2,\  \  \
\bar{\sigma}^2_{AAPL} =  195.95 \  {(\mbox{bp})}^2.
 $$

\begin{figure}[htbp]
\begin{center}
\includegraphics[height=5.5cm]{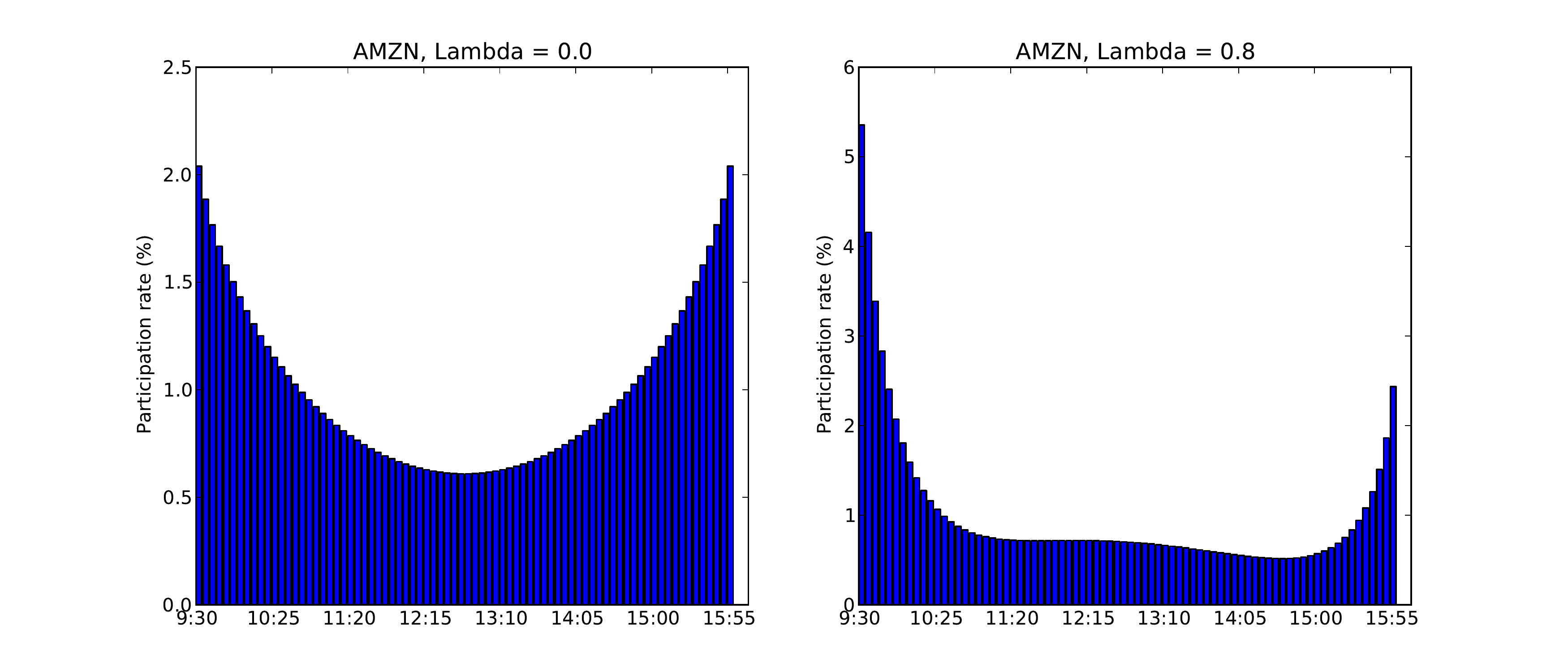}
\includegraphics[height=5.5cm]{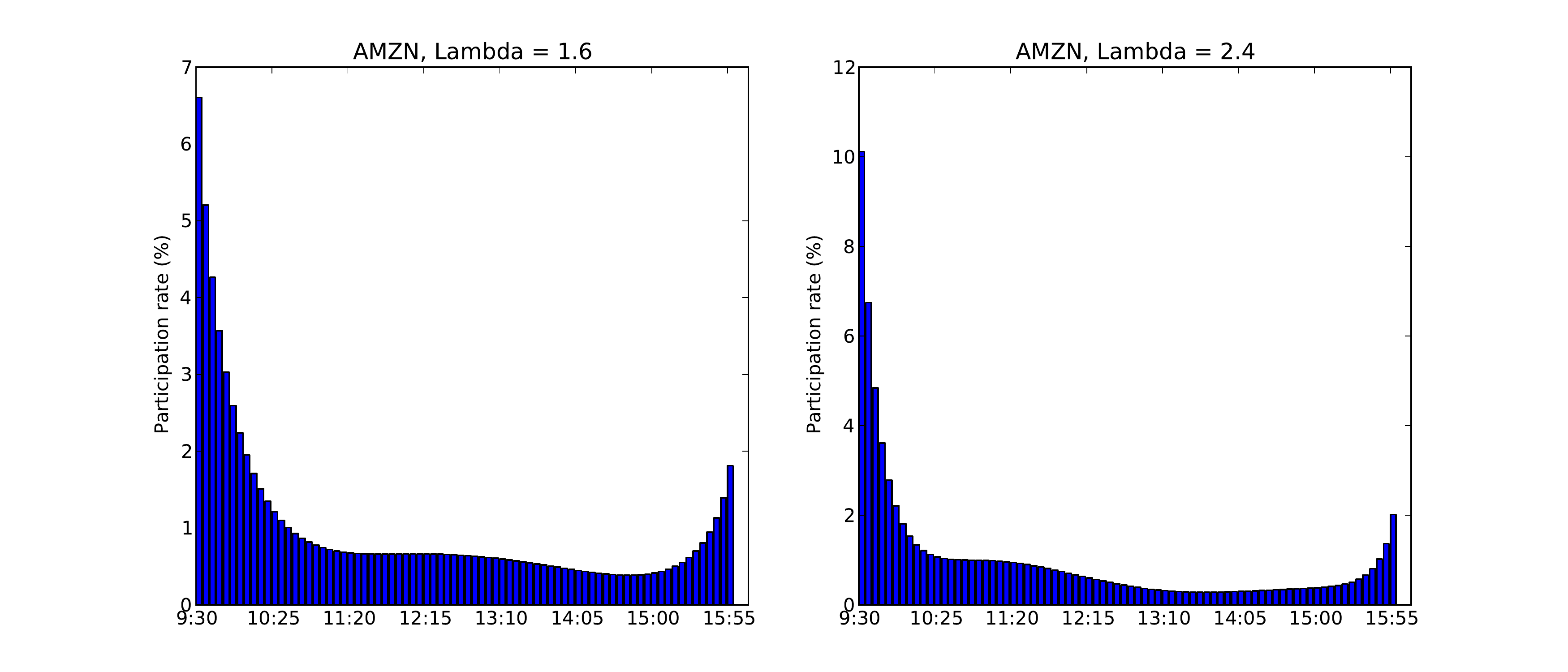}
\end{center}
\caption{Optimal execution strategy to buy 1\% of daily volume of AMZN for four different values of the risk aversion parameter $\lambda$.
}
\label{EmpiricalCalibration_BarcEfficient}
\end{figure}

We apply the same numerical optimization routine we described in the last section, for the minimization of the objective function of Eq. \ref{objFunctDiscrete_chap8}. The result for the stock AMZN is shown in Figure \ref{EmpiricalCalibration_BarcEfficient}. Specifically, we plot the optimal participation rate profile $\mathbf x^*$ for  different values of the risk aversion parameter $\lambda$. As the coefficient $\lambda$ increases, the U-shaped solutions becomes more and more front loaded, so that more trading activity is concentrated at the market opening rather than later in the day. This happens because by trading sooner one is less exposed to the fluctuations of the price and therefore the risk is lower.

As a comparison we consider the optimal  solutions with risk-aversion proposed by Almgren and Chriss \cite{almgren2001optimal} (see Eq. \ref{eq:optimal_solution_alm_chr}).

\begin{figure}[htbp]
\begin{center}
\includegraphics[height=5.5cm]{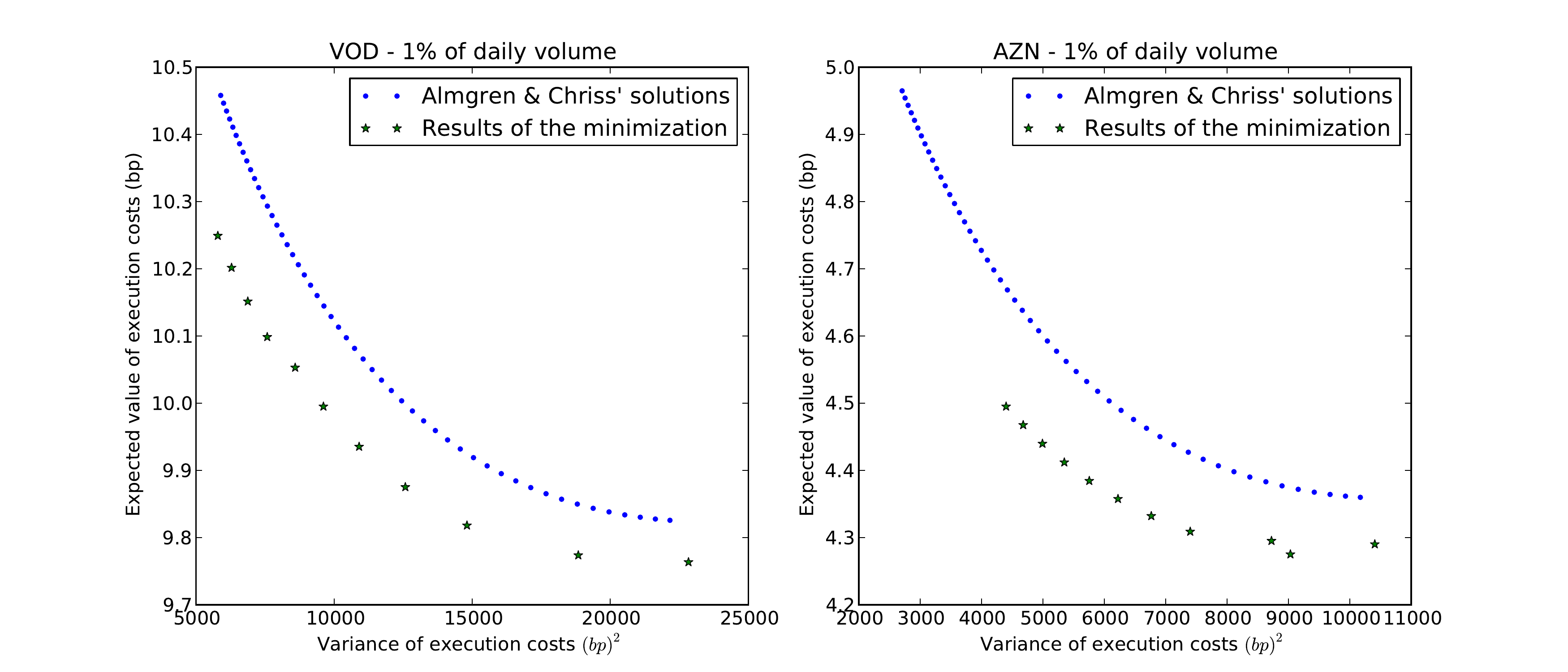}
\includegraphics[height=5.5cm]{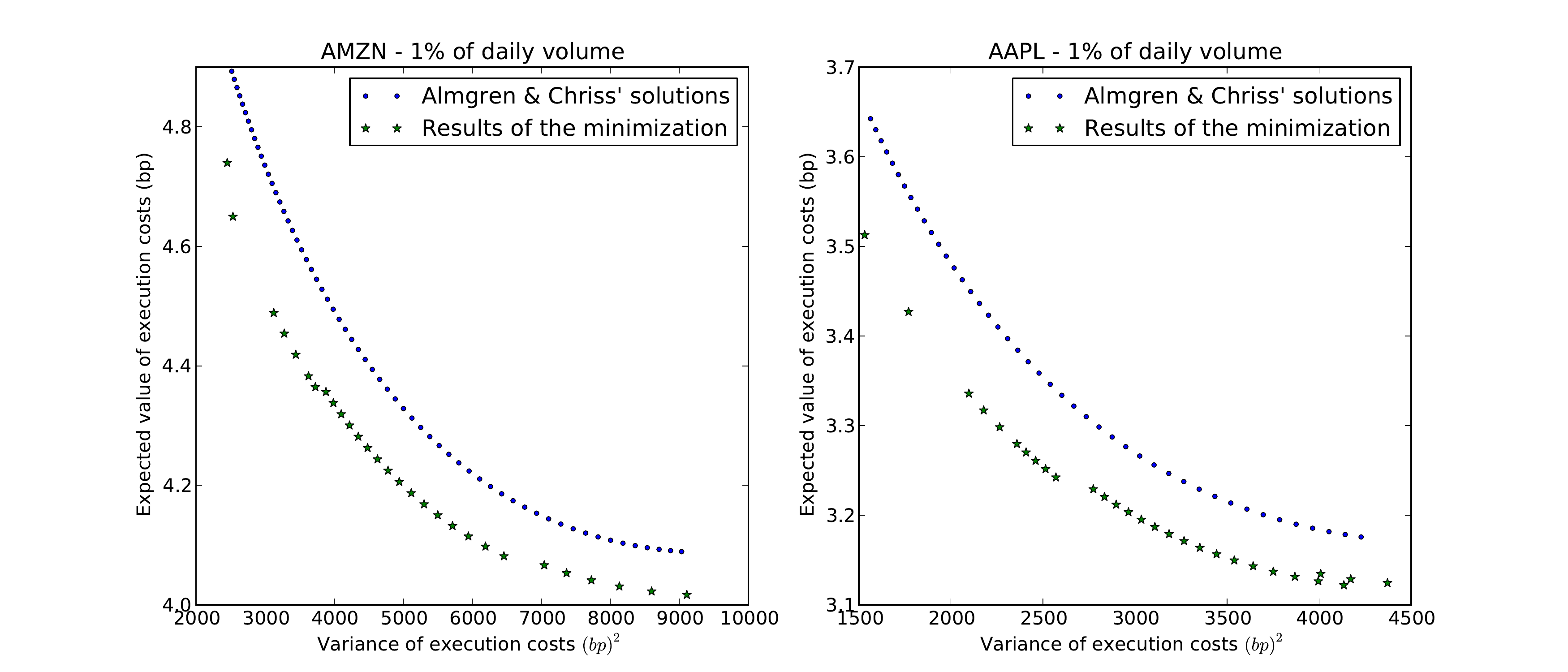}
\end{center}
\caption{Optimal frontier to buy 1\% of daily volume, for the four considered stocks.
Stars correspond to the frontier obtained by using the propagator impact model, while dots refer to the frontier for the Almgren and Chriss model.
On the $x$ axis we have the variance of the execution costs, and on the $y$ axis the expected value of the execution costs (per share).}
\label{EmpiricalCalibration_OptimalFrontier_NEWOptimalFrontier}
\end{figure}

In the case $\lambda=0$, Almgren and Chriss' solution is flat as in the Bertismas Lo model.  We already know that the $\lambda=0$ optimal solutions of propagator model outperforms the flat solution by 1-2\% in impact costs. In order to compare the two models when $\lambda>0$, we plot the efficient frontier of the optimal execution according to the two models. Specifically, for each value of $\lambda$ we compute the expected cost and its variance and we plot them on a cartesian plane. The result is shown in figure \ref{EmpiricalCalibration_OptimalFrontier_NEWOptimalFrontier}.
The dots correspond to Almgren and Chriss' solutions, the stars to the solutions resulting from our minimization procedure. We see that our optimal solutions outperform the $v^{AC}_k$ solutions on the whole range of parameters of risk aversion $\lambda$. For any given level of the variance of the cost, our optimal solution has lower expected impact costs than the corresponding $v^{AC}_k$ of an amount that is roughly constant (at about 1-2\%). It is important to stress however that this is an in sample exercise and therefore the results might be quite different if one considers out of sample real executions. This comparison serves mainly to give an idea of the difference in the results obtained with the two execution strategies.

\section{Conclusions}

In conclusion we have presented an in depth analysis of the way in which the transient impact model introduced and discussed in Refs. \cite{bouchaud2004fluctuations,bouchaud2009markets} to describe the relation between order flow and price at the tick by tick level can be extended to describe impact on a longer time scale. We have considered (i) real time and (ii) aggregated trade time, where time advances by one unit any time a predetermined number of trades is executed in the market (see Appendix). We have shown how to calibrate the model and our empirical analysis has shown that the model fits quite well real data in both settings. This good agreement between model and data opens up the possibility of using the transient impact model for the optimal execution of large orders. By focusing on the real time setting, we have shown how the the model can be used to calibrate optimal execution in different cases. Specifically we have considered the role of risk aversion (following the approach of Ref. \cite{almgren2001optimal}), and of bid ask spread costs. Interestingly, when spread costs are neglected the solution turns out to be composed of alternating buy and sell trading intervals, irrespectively of the position one wants to take with the trade. The introduction of bid ask spread costs regularizes the solution, i.e. one always buys for buy trades. Numerical in sample calculations show that the proposed execution algorithm outperform existing algorithms. 
\section{Appendix: propagator model in aggregated trade time}
\label{subsec:aggreg_trade_time}

In transaction by transaction time we increase time by one unit after each transaction. In aggregated trade time we aggregate $d \in \mathds{N}$ transactions, so that time is increased by one unit after $d$ transactions on the market.

Let us denote $t^{att}_0$  the time of the first transaction of the period we consider
(``$att$'' stands for aggregated trade time).
Then, every $d$ transactions we increase time by one unit and we have that $t^{att}_n$ is the time of the $(n\cdot d)$-th transaction.
The stock price $p^{att}_n$ is defined as the log-midprice right before the $(n\cdot d)$-th transaction. The volume $v^{att}_n$ is the algebraic sum of volumes of the $d$ transactions between time $t^{att}_n$ and time $t^{att}_{n+1}$.
We can express the volume $v^{att}_n$ in term of single trades volumes $v^{tt}_i$ ($tt$ stands for transaction by transaction) as

\begin{equation}
  v^{att}_n =  \sum_{i = d \cdot n}^ {d \cdot (n+1) - 1} v^{tt}_i.
\end{equation}
Clearly for  $d = 1$ we obtain again the transaction by transaction time. 
 
\subsection{Calibration of the propagator model}

We start by estimating the  impact function $f(v^{att})$. This is the expected value of the log mid price change in an interval of $d$ trades conditional to a given volume imbalance $v^{att}$. This quantity has been studied in many papers on market impact (see Ref. \cite{bouchaud2009markets} and references therein).

\begin{figure}[htbp]
\begin{center}
\includegraphics[height=5.5cm]{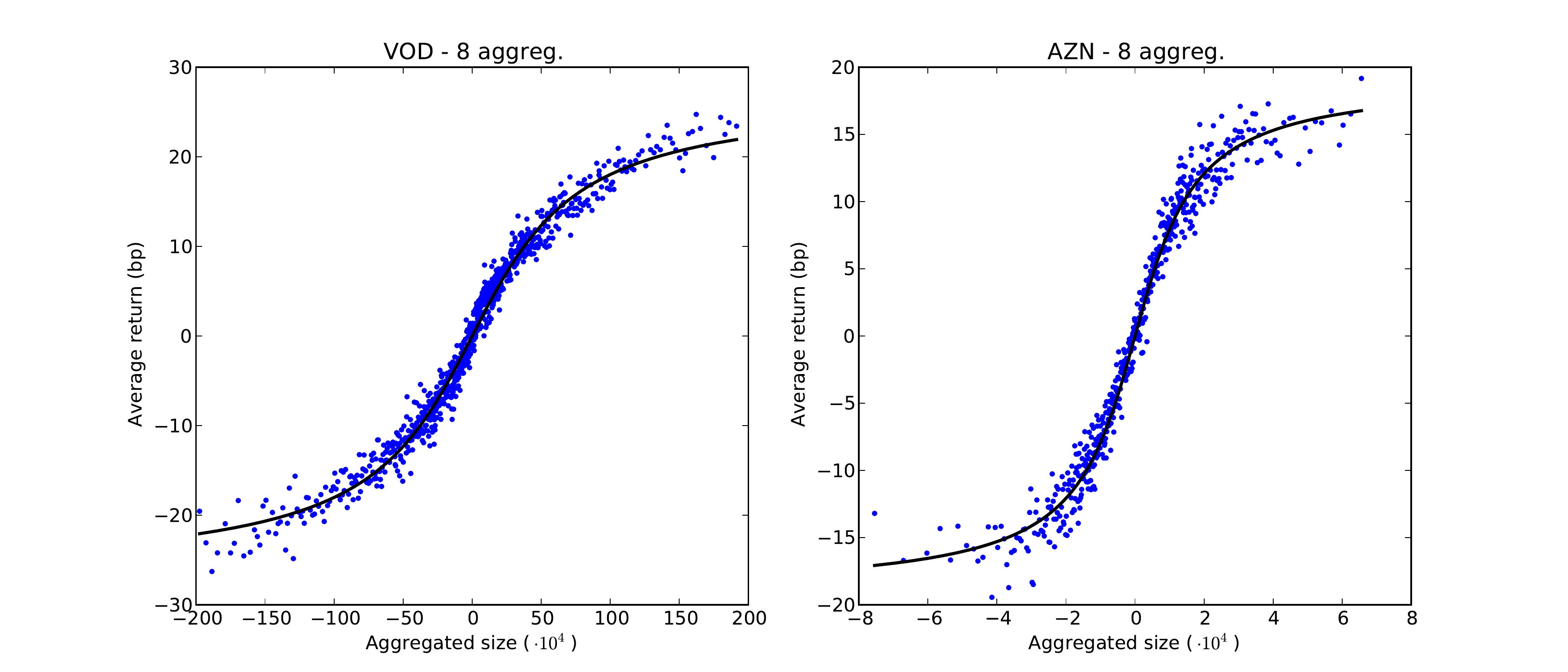}
\includegraphics[height=5.5cm]{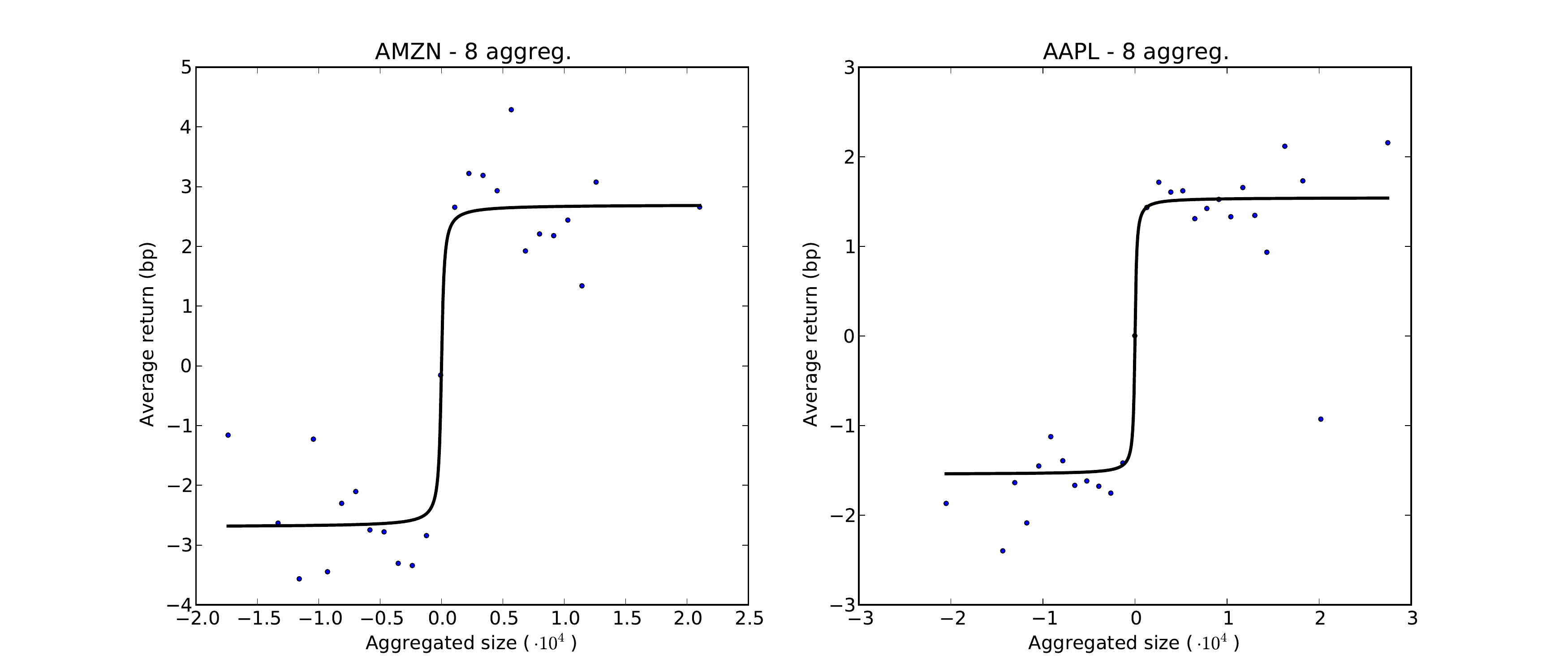}
\end{center}
\caption{Impact of 8 aggregated transactions for the four considered stocks.
}
\label{VolumeImpactAggregTrans_8Lag_ImpactFitted}
\end{figure}

We show in figure \ref{VolumeImpactAggregTrans_8Lag_ImpactFitted} the plots of $f(v^{att})$ in aggregated trade time when we aggregate  $d = 8$ trades for the two sets.
We note that for the long LSE dataset the function is close to be linear for small volumes $v^{att}$, and reaches a constant value for high values of  $v^{att}$. For the NASDAQ stocks the noise level is very high but a clear saturation of the impact for large volumes emerges.
We therefore fit the impact with the arctangent function, as
\begin{equation}
  f(v^{att}) =\theta \arctan(\rho \   v^{att}).
  \label{eq:impact_function_att}
\end{equation}

\begin{figure}[htbp]
\begin{center}
\includegraphics[height=5.5cm]{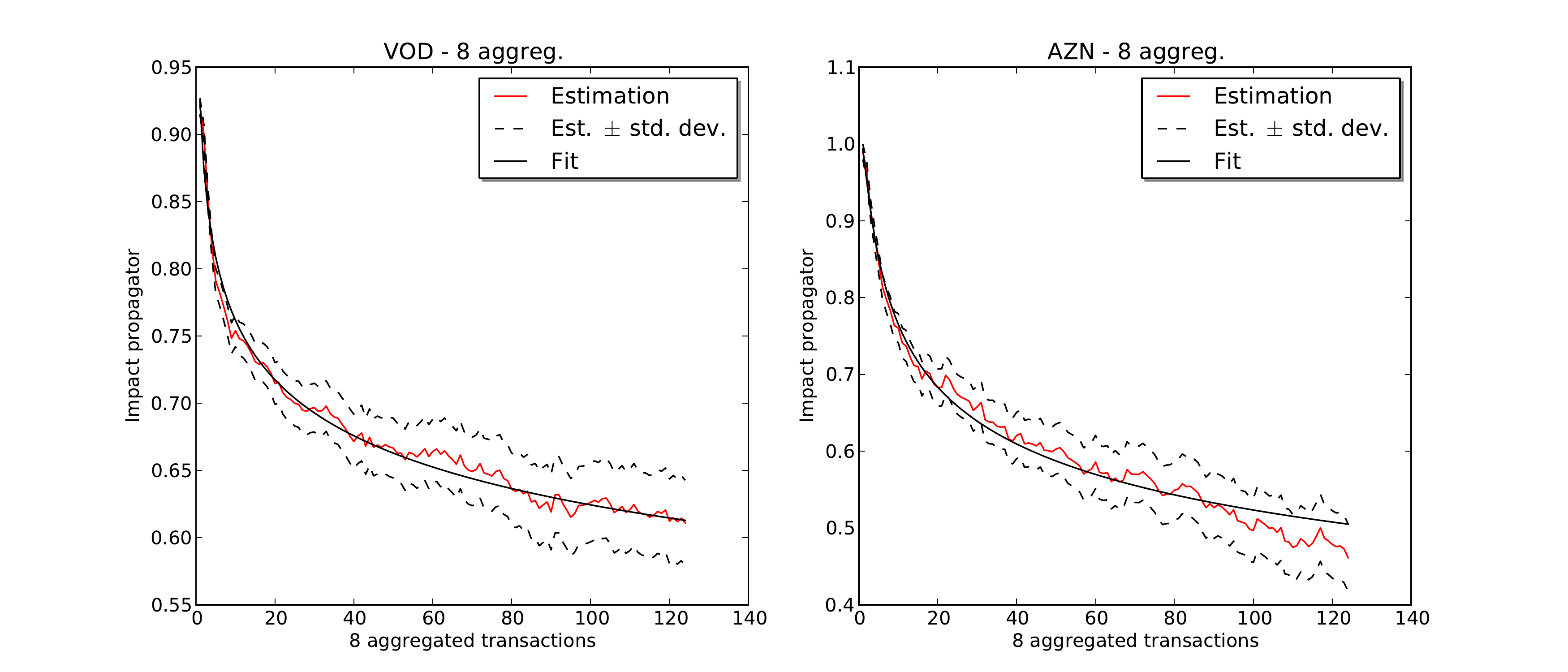}
\includegraphics[height=5.5cm]{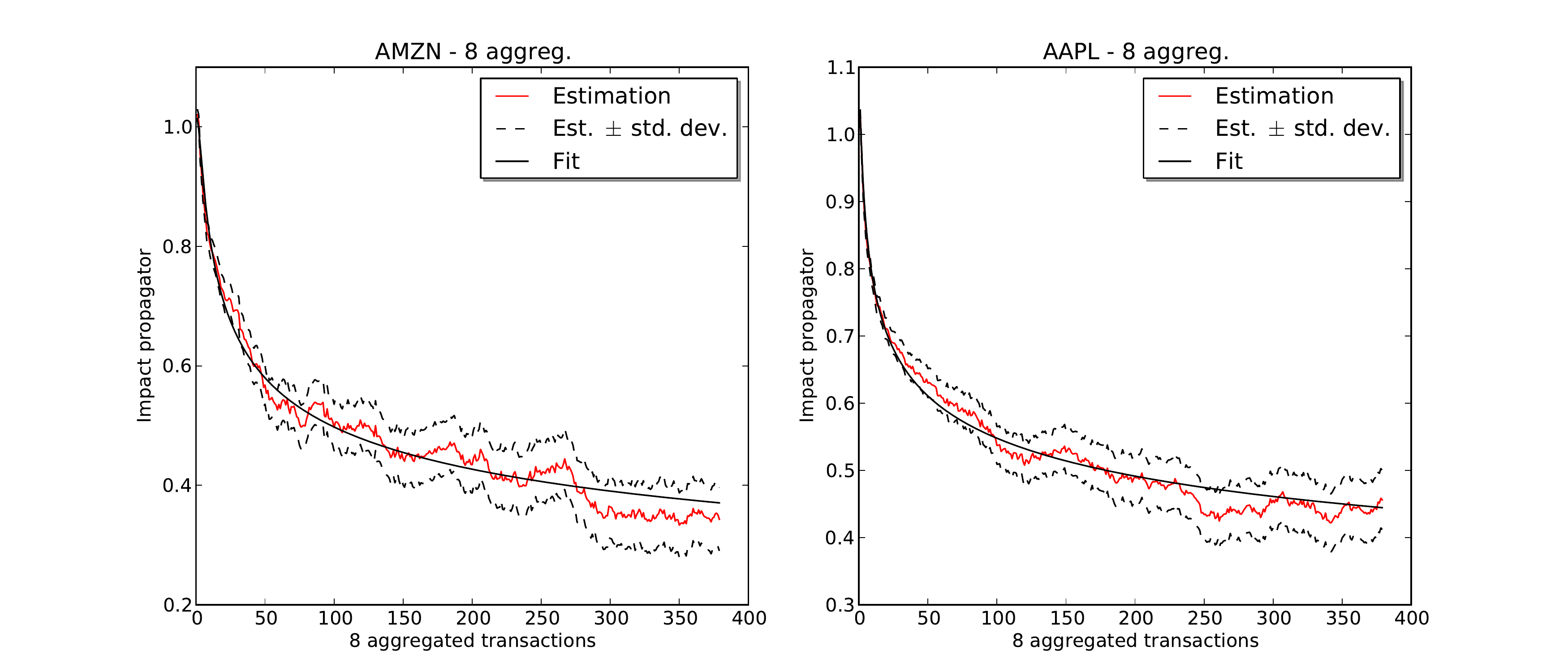}
\end{center}
\caption{Estimation of the impact propagator $G_0$ in aggregated time of 8 transactions for the four considered stocks.
}
\label{G_0_AggregTrans_8Lag_G_0NEW_fitted}
\end{figure}

We can now estimate the impact propagator $G_0$ in aggregated trade time, mimicking the procedure used in real time.
The results  are shown in figure \ref{G_0_AggregTrans_8Lag_G_0NEW_fitted}.
We also fitted a functional form for the impact propagator $G_0$ as in eq. \ref{BouchaudProposedG_0}. 
We note that the propagator model seems to fit quite well the impact decay in aggregated trade time, even if we point out that there is no theoretical reason of why the propagator $G_0$ should preserve its functional shape under aggregation. To the best of our knowledge the theoretical problem of the description of the propagator model under aggregation of transaction is still open.

\bibliographystyle{unsrt}
\bibliography{biblio}

\begin{thebibliography}{10}

\bibitem{bouchaud2004fluctuations}
J.P. Bouchaud, Y.~Gefen, M.~Potters, and M.~Wyart.
\newblock Fluctuations and response in financial markets: the subtle nature of
  `random'price changes.
\newblock {\em Quant. Financ.}, 4(2):176--190, 2004.

\bibitem{kissell2003optimal}
R.~Kissell, M.~Glantz, and R.~Malamut.
\newblock {\em Optimal Trading Strategies: Quantitative Approaches for Managing
  Market Impact and Trading Risk}.
\newblock AMACOM, New York, NY, 2003.

\bibitem{bertsimas1998optimal}
D.~Bertsimas and A.~Lo.
\newblock Optimal control of execution costs.
\newblock {\em J. Financ. Mark.}, 1(1):1--50, 1998.

\bibitem{almgren2001optimal}
R.~Almgren and N.~Chriss.
\newblock Optimal execution of portfolio transactions.
\newblock {\em J. Risk}, 3:5--40, 2001.

\bibitem{bouchaud2009markets}
J.P. Bouchaud, J.D. Farmer, and F.~Lillo.
\newblock {\em How markets slowly digest changes in supply and demand},
  volume~4 of {\em Handbook of financial markets}.
\newblock North-Holland, San Diego, CA, 2009.

\bibitem{hopman2002supply}
C.~Hopman.
\newblock Are supply and demand driving stock prices?
\newblock {\em Quant. Financ.}, 12:104--145, 2002.

\bibitem{lillo2003master}
F.~Lillo, J.D. Farmer, and R.N. Mantegna.
\newblock Master curve for price-impact function.
\newblock {\em Nature}, 421(129):176--190, 2003.

\bibitem{farmer2004really}
J.D. Farmer, L.~Gillemot, F.~Lillo, S.~Mike, and A.~Sen.
\newblock What really causes large price changes?
\newblock {\em Quant. Financ.}, 4:383--397, 2004.

\bibitem{potters2003more}
M.~Potters and J.P. Bouchaud.
\newblock More statistical properties of order books and price impact.
\newblock {\em Physica A}, 324(1-2):133--140, 2003.

\bibitem{lillo2004long}
F.~Lillo and J.D. Farmer.
\newblock The long memory of the efficient market.
\newblock {\em Stud. Nonlinear Dyn. E.}, 8(3):1--32, 2004.

\bibitem{imon}
F.~Toth B.~Palit, I.~Lillo and Farmer J.D.
\newblock Why is order flow so persistent?
\newblock Preprint, arXiv:1108.1632, 2011.

\bibitem{alfonsi2009order}
A.~Alfonsi, A.~Schied, and A.~Slynko.
\newblock Order book resilience, price manipulation, and the positive portfolio
  problem.
\newblock Preprint, arXiv:0708.1756v3, 2009.

\bibitem{obizhaeva2006optimal}
Anna Obizhaeva and Jiang Wang.
\newblock Optimal trading strategy and supply/demand dynamics.
\newblock NBER Working Papers, http://ideas.repec.org/p/nbr/nberwo/11444.html
  11444, National Bureau of Economic Research, June 2005.

\bibitem{perold1998implementation}
A.F. Perold.
\newblock The implementation shortfall: Paper versus reality.
\newblock {\em J. Portfolio Manage.}, 31:106, 1998.

\bibitem{markowitz1968portfolio}
H.M. Markowitz.
\newblock {\em Portfolio selection: Efficient diversification of investments}.
\newblock Yale University Press, New Haven, CT, 1968.

\bibitem{elton1995persistence}
E.~Elton, M.~Gruber, and C.~Blake.
\newblock The persistence of risk-adjusted mutual fund performance.
\newblock NYU, Working Paper. Available at SSRN:
  http://ssrn.com/abstract=1298325, 1995.

\bibitem{lee1991inferring}
C.M.C. Lee and M.J. Ready.
\newblock Inferring trade direction from intraday data.
\newblock {\em J. Financ.}, 46:733--746, 1991.

\bibitem{bouchaud2006random}
J.P. Bouchaud, J.~Kockelkoren, and M.~Potters.
\newblock Random walks, liquidity molasses and critical response in financial
  markets.
\newblock {\em Quant. Financ.}, 6(2):115--123, 2006.

\bibitem{gatheral2009no}
J.~Gatheral.
\newblock No-dynamic-arbitrage and market impact.
\newblock {\em Quant. Financ.}, 10(7):749--759, 2009.

\bibitem{byrd1995limited}
R.H. Byrd, P.~Lu, J.~Nocedal, and C.~Zhu.
\newblock A limited memory algorithm for bound constrained optimization.
\newblock {\em J. Sci. Comput.}, 16(5):1190--1208, 1995.

\bibitem{zhu1997algorithm}
C.~Zhu, R.H. Byrd, P.~Lu, and J.~Nocedal.
\newblock Algorithm 778: L-bfgs-b: Fortran subroutines for large-scale
  bound-constrained optimization.
\newblock {\em ACM T. Math. Software}, 23(4):550--560, 1997.

\bibitem{Scipy}
E.~Jones, T.~Oliphant, and P.~et~al. Peterson.
\newblock {SciPy}: Open source scientific tools for {Python}, 2001--.

\bibitem{L1pnas}
J.~Brodie, I.~Daubechies, C.~De Mol, D.~Giannone, and I.~Loris.
\newblock Sparse and stable markowitz portfolios.
\newblock {\em Proceedings of the National Academy of Sciences},
  106:12267--12272, 2009.

\bibitem{caccioli11}
F.~Caccioli, S.~Still, M.~Marsili, and I.~Kondor.
\newblock Optimal liquidation strategies regularize portfolio selection.
\newblock {\em The European Journal of Finance}, 1:1--18, 2011.

\end{thebibliography}

\clearpage

\section*{Disclaimer}
{\footnotesize
This material ({\bf Material}) is provided for your information only and does not constitute: (i) research or a product of the J.P. Morgan research department, (ii) an offer to sell, a solicitation of an offer to buy, or a recommendation for any investment product or strategy, or (iii) any investment, legal or tax advice. You are solely responsible for deciding whether any investment product or strategy is appropriate for you based upon your investment goals, financial situation and tolerance for risk. J.P. Morgan disclaims all representations and warranties in the information contained in this Material, whether express or implied, including, without limitation, any warranty of satisfactory quality, completeness, accuracy, fitness for a particular purpose or non-infringement.  The information contained herein is as of the date and time referenced in the Material and J.P. Morgan does not undertake any obligation to update such information.  All content, data, statements and other information are not warranted as to completeness or accuracy and are subject to change without notice.  J.P. Morgan disclaims any responsibility or liability, whether in contract, tort (including, without limitation, negligence), equity or otherwise, for the quality, accuracy or completeness of the information contained in this Material, and for any reliance on, or uses to which, this Material, is put, and you are solely responsible for any use to which you put such information. Without limiting any of the foregoing, to the fullest extent permitted by applicable law, in no event shall J.P. Morgan have any liability for any special, punitive, indirect, or consequential damages (including lost profits or lost opportunity), in connection with the information contained in this Material, even if notified of the possibility of such damages. Any comments or statements made herein do not necessarily reflect those of J.P. Morgan, its subsidiaries or its affiliates. Any unauthorized use, dissemination, distribution or copying of this Material, in whole or in part, is strictly prohibited. \\
\copyright 	\   2012 JP Morgan Chase \& Co.  All rights reserved.  J.P. Morgan is the global brand name for J.P. Morgan Chase \& Co. and its subsidiaries and affiliates worldwide. J.P. Morgan Cazenove is a marketing name for the U.K. investment banking businesses and EMEA cash equities and equity research businesses of JPMorgan Chase \& Co. and its subsidiaries, conducted primarily through J.P. Morgan Securities Ltd. (ÒJPMSLÓ).  Execution services are offered through J.P. Morgan Securities LLC ( member of FINRA, NYSE and SIPC),  JPMSL ( member of the London Stock Exchange and is authorized and regulated by the Financial Services Authority. Registered in England \& Wales No. 2711006. Registered Office 125 London Wall, London EC2Y 5AJ), J.P. Morgan Securities (Asia Pacific) Limited (CE number AAJ321) ( regulated by the Hong Kong Monetary Authority and the Securities and Futures Commission in Hong Kong), and other investment banking affiliates and subsidiaries of JP Morgan in other jurisdictions worldwide and registered with local authorities as appropriate.  Please consult http://www.jpmorgan.com/pages/jpmorgan/investbk/global for more information.  Prime brokerage services, including clearing and custody services, are offered by J.P. Morgan Clearing Corp. (member of FINRA, NYSE and SIPC).  Product names, company names, and logos mentioned herein are trademarks or registered trademarks of their respective owners.
}
  \end{document}